\documentclass[aps,prd,reprint,superscriptaddress,floatfix]{revtex4-2}

\usepackage{amsmath,amssymb,amsfonts}
\usepackage{bm}
\usepackage{graphicx}
\usepackage{dcolumn}
\usepackage{booktabs}

\usepackage[dvipsnames]{xcolor}

\definecolor{PRDBlue}{HTML}{174A77}
\definecolor{PRDTeal}{HTML}{087E8B}
\definecolor{PRDViolet}{HTML}{69418C}

\usepackage{hyperref}
\hypersetup{
    colorlinks=true,
    linkcolor=PRDBlue,
    citecolor=PRDTeal,
    urlcolor=PRDViolet
}

\begin{document}

\title{Landauer bound in a regular Bardeen black hole: extremality and evaporation}

\author{J. Andrade}
\email{julio.andrade@espoch.edu.ec}
\affiliation{Facultad de Ciencias, Escuela Superior Polit\'ecnica de Chimborazo (ESPOCH), Riobamba 060155, Ecuador}

\author{J. O. Cede\~no}
\affiliation{Centro de Investigaci\'on y Desarrollo Cient\'ifico Z\'etesis, Riobamba, Ecuador}
\affiliation{Facultad de Ciencias, Escuela Superior Polit\'ecnica de Chimborazo (ESPOCH), Riobamba 060155, Ecuador}

\author{Ji\v{r}\'i Svozil\'ik}
\affiliation{Facultad de Ciencias, Escuela Superior Polit\'ecnica de Chimborazo (ESPOCH), Riobamba 060155, Ecuador}

\author{Tayyab Naseer}
\email{tayyabnaseer48@yahoo.com}
\affiliation{Department of Mathematics and Statistics, The University of Lahore, 1-KM Defence Road, Lahore-54000, Pakistan}
\affiliation{Jadara Research Center, Jadara University, Irbid 21110, Jordan}

\author{Faisal Javed}
\affiliation{Department of Physics, Zhejiang Normal University, Jinhua 321004, China}
\affiliation{Department of Mathematics, School of Science, University of Management \\and Technology, Lahore 54000, Pakistan}

\author{M. Zeeshan Gul}
\email{mzeeshangul.math@gmail.com (corresponding author)}
\affiliation{Department of Mathematics and Statistics, The University of Lahore, 1-KM Defence Road, Lahore-54000, Pakistan}
\affiliation{Research Center of Astrophysics and Cosmology, Khazar University, 41 Mehseti Street, Baku AZ1096, Azerbaijan}


%




\begin{abstract}
Does the vanishing Hawking temperature of an extremal black hole imply that information can be erased without energy transfer? We investigate this question in the regular Bardeen geometry using a two-level quantum system, representing a quantum memory, coupled to a scalar bosonic field confined within an exterior cavity. The Hawking temperature tends to zero at extremality, as does the corresponding thermal Landauer scale. The field in an exterior cavity behaves differently, provided both walls remain away from the horizon, its lowest excitation energy stays finite. For a maximally mixed memory coupled to a reservoir initially in its unique ground state, we find $Q_\infty\geq\Delta_{R,\infty}(1/2-\epsilon)$, with $\Delta_{R,\infty}$ the reservoir gap and $\epsilon$ the reset error. For a resonant single-mode interaction, we give a reset protocol that attains the bound with no work supplied by the external control. In this example, the memory itself supplies the energy gained by the field. In the multimode calculation, counter-rotating terms are retained. Keeping the pulse parameters unchanged, we find a larger residual reset error closer to extremality. For reservoirs prepared thermally, we evaluate the Reeb--Wolf balance and find that the relative-entropy contribution, expressed in energy units, can have a finite low-temperature limit. These numerical preparations are externally imposed and remain hotter than the Hawking temperature. For comparison, an unconfined scalar field has near-extremal luminosity $P_\infty\propto y^4$, with $y=r_H/|g|-\sqrt2$. Assuming a particular quasistatic energy-loss prescription gives $y\propto t^{-1/2}$; this does not establish the existence of a stable remnant. For this particular reset operation, vanishing Hawking temperature therefore does not force the transferred heat to vanish.\\


\end{abstract}

\maketitle

\section{Introduction}

Landauer's principle states that the erasure of information in an irreversible logical process has significant thermodynamic consequences. Specifically, a logically irreversible erasure that reduces the entropy of a memory is accompanied by a transfer of heat to the environment and an increase in the entropy of the surroundings \cite{landauer1961irreversibility}. This principle is consistent with the second law of thermodynamics. For an initially equiprobable bit erased in contact with a reservoir at temperature $T$, the heat deposited into the reservoir satisfies
\begin{equation}\label{Q_min}
 Q \geq Q_{\min}= k_{B}T\ln2.
\end{equation}
Here, $T$ is the reservoir temperature and $k_B$ denotes Boltzmann's constant. This has been experimentally verified in several studies related to information erasure in one-bit memories, including designs involving colloidal particles in bistable potentials and nanomagnetic memory elements \cite{berut2012experimental, jun2014high, hong2016experimental, gaudenzi2018quantum, yan2018single}.

Subsequently, Reeb and Wolf \cite{reeb2014improved} provided a rigorous quantum formulation of Landauer's principle for finite reservoirs, proposing that the heat received by the reservoir decomposes into contributions associated with the system's entropy reduction, the correlations generated between the memory and the reservoir, and the reservoir's deviation from its initial thermal state. For an initially uncorrelated memory and a thermal reservoir undergoing joint unitary evolution, their equality relates the reservoir heat to the memory entropy decrease, the final mutual information, and the reservoir relative entropy. The balance used in our calculation is developed in Sec.~\ref{sec:balance}.

In a gravitational setting, the temperature relevant to erasure must be distinguished from the location at which the exchanged energy is evaluated. Fuchs \cite{fuchs1992landauer} linked information erasure to black hole entropy, while Daffertshofer and Plastino \cite{daffertshofer2007forgetting} examined erasure in a gravitational field and its connection to the Tolman temperature. Song and Winstanley \cite{song2008information} analysed information erasure in connection with the generalized second law, and Kim, Lee, and Lee \cite{kim2010black} studied the interplay between black hole thermodynamics and information loss. Herrera considered erasure at black hole horizons \cite{herrera2008possible} and subsequently formulated Landauer's principle explicitly within general relativity \cite{herrera2020landauer}.

Specifically, L. Herrera notes that the erasure of a bit of information must take into account the action of the gravitational field in which the system is embedded; the process must therefore be modified as follows
\begin{equation}\label{Q_min_gtt}
   Q_{\infty} \geq k_{B}T_{\mathrm{loc}}N(r)\ln2,
\end{equation}
where $N(r)$ is the dimensionless lapse normalized to unity at infinity and $T_{\mathrm{loc}}$ denotes the local temperature. For the metric used below, $N(r)=\sqrt{f(r)}$; with $t$ measured in seconds, $g_{tt}=-c^2f(r)$ and $N(r)=\sqrt{|g_{tt}|}/c$. For a static spacetime with the time coordinate normalized at infinity, the Tolman relation $T_{\infty}=T_{\mathrm{loc}}N(r)$ identifies the redshifted temperature. Consequently, Eq. (\ref{Q_min_gtt}) must be interpreted as the minimum erasure energy referred to infinity; the local Landauer limit retains the standard form $Q_{\min}^{\mathrm{loc}}\geq k_{B}T_{\mathrm{loc}}\ln 2$. This distinction is important because the lapse represents the gravitational redshift, rather than a modification of the local thermodynamic statement itself.

More recently, the connection between information erasure and black hole thermodynamics has developed in several directions. Cort\^{e}s and Liddle \cite{cortes2024hawking} argued that Hawking evaporation saturates the Landauer limit, thereby relating black hole mass loss to the thermodynamic cost of information processing. Abreu and Ananias Neto \cite{abreu2025modified} investigated how generalized entropy prescriptions modify information-erasure bounds and black hole mass loss, whereas Kemple \cite{kemple2026landauer} estimated the maximum number of irreversible bit operations available to a Schwarzschild black hole. Ananias Neto and Thibes \cite{neto2026generalized} connected Landauer's principle with horizon-area quantisation in generalized-entropy frameworks. More recently, Andrade and collaborators \cite{andrade2026reeb} extended the analysis to ``hairy'' black holes generated via gravitational decoupling, employing the Reeb--Wolf formulation to isolate mutual-information and relative-entropy contributions to the erasure cost and to analyse the influence of gravitational hair in both continuous and area-quantized regimes.

These developments raise a concrete question for regular black holes does the vanishing of $k_BT_H\ln2$ at extremality also remove the energy transfer required by an erasure operation? The Bardeen geometry \cite{bardeen1968non, ayon2000bardeen} allows this question to be studied along a family whose Hawking temperature tends to zero while the horizon area remains finite. We place a two-level quantum system and a scalar-field cavity outside the horizon and follow how the cavity spectrum and a fixed reset pulse respond to changes in the geometry. Each operation takes place on a static background.

The spectral restriction on heat used below is not peculiar to Bardeen, a finite reservoir gap can constrain erasure in other settings. The contribution of this calculation is to determine that gap, the gravitational redshift, and the mode couplings for a specified cavity in the Bardeen exterior, and to connect them to the reset error and the Reeb--Wolf balance. This makes it possible to compare a vanishing Hawking thermal scale with a protocol whose accessible excitation energies remain finite. We then examine unconfined Hawking emission on the same family. That separate calculation addresses whether the suppression of radiative power changes the interpretation of the erasure cost; it does not supply a thermal bath for the cavity. Its evaporation consequences are considered only within an explicitly conditional quasistatic prescription.

\section{Regular Bardeen geometry}
\label{sec:framework}
We consider the static and spherically symmetric line element
\begin{equation}
    ds^{2} = -f(r)c^2dt^2 + f(r)^{-1}dr^2 + r^{2}d\Omega_{2}^2,
\end{equation}
with $d\Omega_{2}^{2} = d\theta^{2}+\sin^{2}\theta\,d\phi^{2}$, and
\begin{equation}
    f(r) = 1-\frac{2m r^2}{(r^2+g^2)^{3/2}},
\end{equation}
where $m$ is the gravitational length associated with the ADM mass $\mathcal{M}$, given by
\begin{equation}
    m = \frac{G\mathcal{M}}{c^2},
\end{equation}
and $g$ denotes the magnetic-charge parameter, with dimensions of length, \cite{bardeen1968non}. This metric describes a regular black-hole geometry. 

\begin{figure*}[tbp]
\centering
\includegraphics[width=0.47\textwidth]{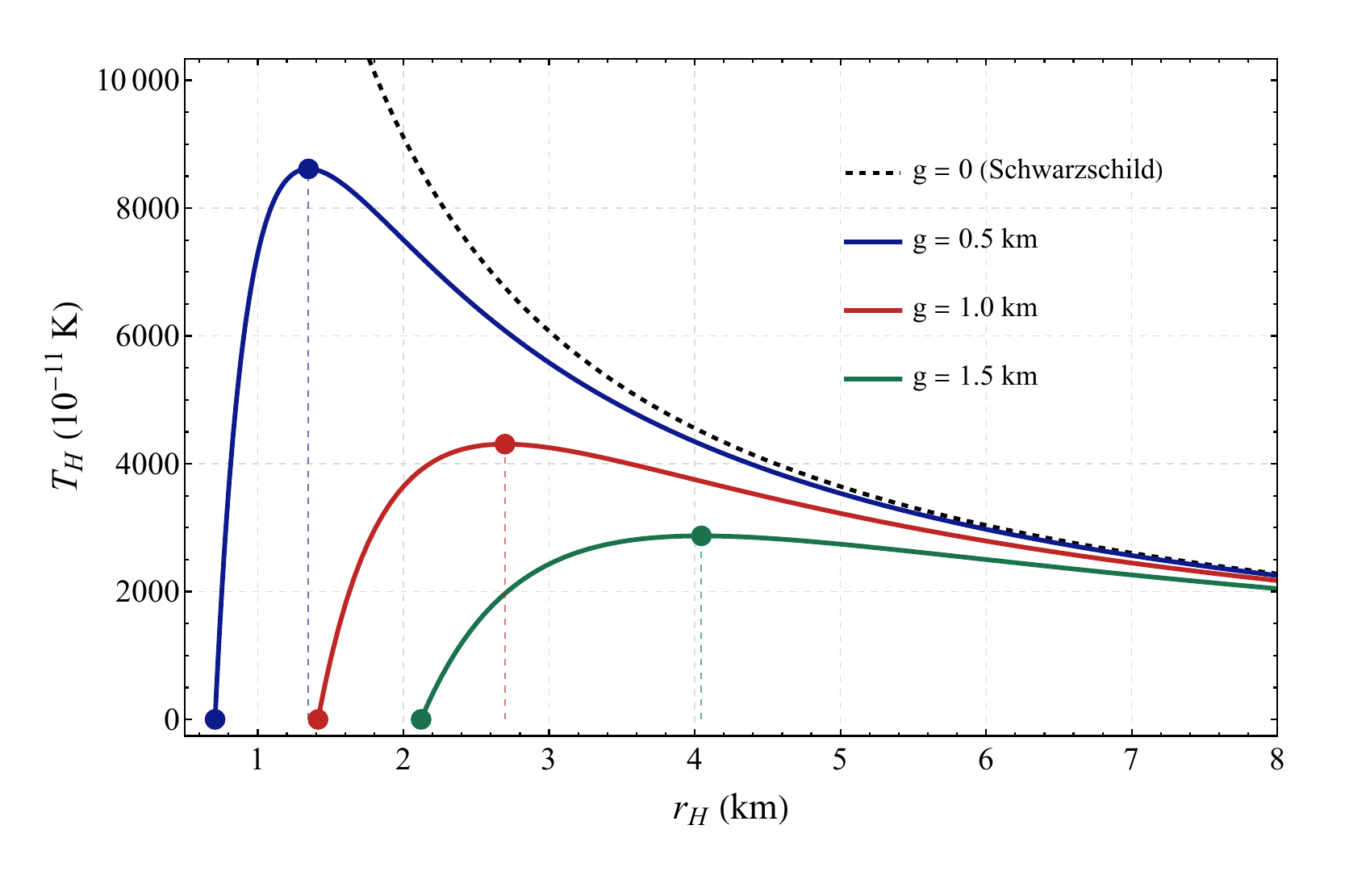}\hfill
\includegraphics[width=0.47\textwidth]{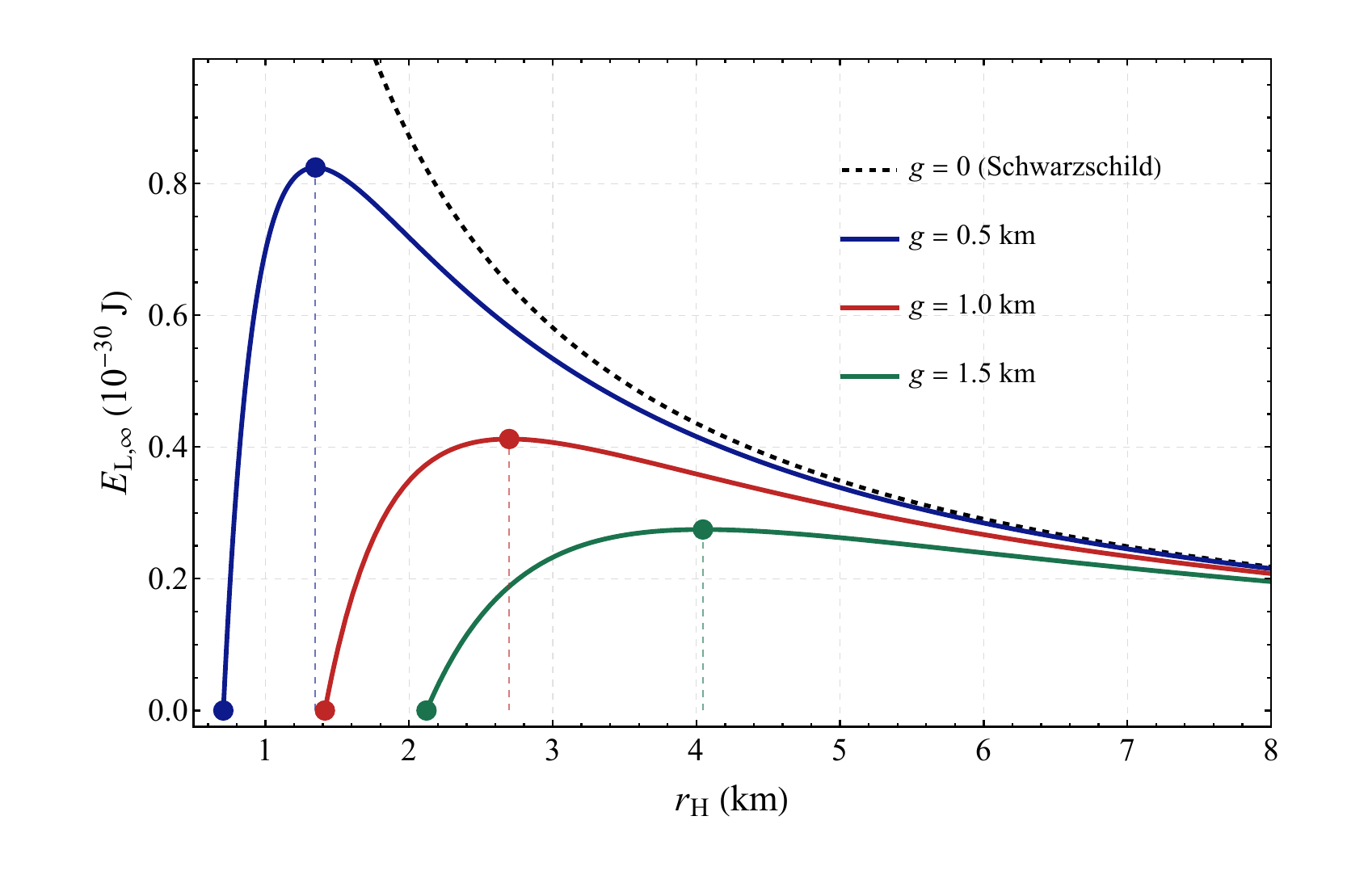}
\caption{Hawking temperature (left) and one-bit thermal Landauer scale (right) as functions of the outer-horizon radius. The two panels are related exactly by $E_{L,\infty}=k_B\ln2\,T_H$. Points mark the extremal configurations and maxima for $g>0$. The dashed Schwarzschild reference extends above the plotted range at small radii.}
\label{fig:1}\label{fig:2}
\end{figure*}

The horizon radii are determined by the positive roots of
\begin{equation}
    f(r_{H}) = 0,
\end{equation}
which gives
\begin{equation}\label{mass_m}
    m = \frac{(r_{H}^2+g^2)^{3/2}}{2r_{H}^2}.
\end{equation}
To determine the extremal horizon and mass, we impose the double-root conditions 
\begin{equation}
    f(r_{e}) = 0, 
\end{equation}
and
\begin{equation}\label{fdev}
    f'(r_{e}) = 0.
\end{equation}
For $m>0$ and $g\neq 0$, these conditions imply
\begin{equation}
    r_{e} = \sqrt{2}\vert g\vert,
\end{equation}
while the extremal gravitational mass parameter follows from Eq.~(\ref{mass_m}) as
\begin{equation}
    m_{e} =  \frac{3\sqrt{3}}{4}\vert g\vert.
\end{equation}
For the outer horizon, the surface gravity is 
\begin{equation}
    \kappa = \frac{c^2}{2}f'(r_{H}),
\end{equation}
and the corresponding Hawking temperature is 
\begin{equation}\label{T_H}
    T_{H} = \frac{\hbar \kappa}{2\pi k_{B}c} = \frac{\hbar c}{4\pi k_{B}}\frac{r_{H}^{2}-2g^2}{r_{H}(r_{H}^{2}+g^{2})}.
\end{equation}
We restrict the thermal analysis to the outer-horizon branch $r_H\geq\sqrt{2}|g|$, with $g\neq0$. At fixed $g$, the maximum temperature is determined by $dT_H/dr_H=0$, which yields 
\begin{equation}
    r_{max} = \vert g\vert \sqrt{\frac{7+\sqrt{57}}{2}}.   
\end{equation}
Thus, the temperature increases as the horizon radius decreases from large values, reaches a maximum, and subsequently decreases until it vanishes in the extremal configuration (see Fig.~\ref{fig:1}).

At extremality, the surface gravity vanishes
\begin{equation}
    \kappa = \frac{c^2}{2}f'(r_{e}) = 0,
\end{equation}
and therefore $T_{H}\vert_{r = r_{e}}= 0$.

The corresponding one-bit thermal Landauer scale for a reservoir prepared at $T_\infty=T_H$ is
\begin{equation}\label{Emin}
    E_{L,\infty} = \frac{\hbar c \ln 2}{4\pi}\frac{r_{H}^{2}-2g^2}{r_{H}(r_{H}^{2}+g^{2})}.
\end{equation}
Since $E_{L,\infty}=k_B T_H\ln 2$, the two quantities have the same qualitative dependence on $r_H$ (see Fig.~\ref{fig:2}). This is the conventional thermal scale for ideal one-bit erasure; it does not, however, determine the heat released in a particular operation. Nor does it establish whether a finite reservoir can reset a bit perfectly.

\section{Horizon entropy and thermodynamic consistency}

The fact that the Hawking temperature vanishes at extremality raises an important question about the behavior of black hole entropy. In Einstein gravity, the entropy associated with the event horizon is given by
\begin{equation}\label{entropy_BH}
    S_{BH} = \frac{\pi k_{B} r_{H}^{2}}{\ell_{P}^{2}},
\end{equation}
where $\ell_{P} = \sqrt{\frac{G\hbar}{c^{3}}}$ is the Planck length. For the extremal Bardeen configuration, this gives the nonzero entropy
\begin{equation}
    S_{e} = \frac{2\pi k_{B} g^{2}}{\ell_{P}^{2}}.
\end{equation}
The extremal member of the static Bardeen family has zero Hawking temperature and finite geometric entropy. This statement does not establish that a dynamical evaporation process reaches extremality in finite time.

This feature is particularly relevant because, in the extremal limit, Landauer's thermal energy scale vanishes, whereas the geometric entropy of the horizon remains finite and non-zero. However, the vanishing of this thermal scale does not imply that an arbitrary protocol for erasing physical information can be carried out without energy transfer.

Differentiating Eq.~(\ref{mass_m}) at fixed $g$, and using $\mathcal{M}=c^2m/G$, we find
\begin{equation}
 \frac{d\mathcal{M}}{dr_H}=\frac{c^2}{2G}
 \frac{\sqrt{r_H^2+g^2}(r_H^2-2g^2)}{r_H^3}.
\end{equation}
Likewise, Eqs.~(\ref{T_H}) and (\ref{entropy_BH}) imply
\begin{equation}
 T_H\frac{dS_{BH}}{dr_H}=\frac{c^4}{2G}
 \frac{r_H^2-2g^2}{r_H^2+g^2}.
\end{equation}
Consequently, the identity along the fixed-$g$ static family is
\begin{equation}\label{restricted_identity}
\begin{aligned}
 T_H\,dS_{BH}&=\mathcal{C}(r_H,g)c^2\,d\mathcal{M},\\
 \mathcal{C}(r_H,g)&=\frac{r_H^3}{(r_H^2+g^2)^{3/2}}.
\end{aligned}
\end{equation}
It is worth noting that the resulting relationship constitutes a differential identity along a family of static solutions with constant \(g\). It should not be directly interpreted as a complete first law of thermodynamics, since such an interpretation requires considering the contributions of the matter sector that supports the geometry. The role of a mass-dependent source in the first law for regular black holes has been discussed by Ma and Zhao \cite{ma2014corrected}. Here we retain the Einstein-gravity area entropy and use Eq.~(\ref{restricted_identity}) only as an identity along the fixed-$g$ static family.

In particular,
\begin{equation}
    \mathcal{C}_{e} = \left(\frac{2}{3}\right)^{3/2}.
\end{equation}
This identity reveals an important difference compared to the Schwarzschild case. Indeed, when $g\rightarrow 0$ then $\mathcal{C}\rightarrow 1$, and the familiar Schwarzschild relation is recovered
\begin{equation}
    T_{H}dS_{BH} = c^{2}d\mathcal{M}.
\end{equation}

\section{Redshift and the meaning of the thermal bound}

For complete erasure of an initially equiprobable bit at a static radius $r_{0}>r_{H}$, the local heat received by a thermal reservoir obeys
\begin{equation}
  Q_{\mathrm{loc}} \geq k_{B}T_{\mathrm{loc}}\ln 2.
\end{equation}
For a localized energy exchange, the redshift relation between the heat measured locally and that referred to infinity is \cite{herrera2020landauer}
\begin{equation}
 Q_\infty=\sqrt{f(r_0)}\,Q_{\mathrm{loc}}
 \geq k_B T_{\mathrm{loc}}\sqrt{f(r_0)}\ln 2.
\end{equation}
Now, if the reservoir is in thermal equilibrium, then we have that
\begin{equation}
    T_{\mathrm{loc}} \sqrt{f(r_{0})} = T_{\infty}, 
\end{equation}
upon choosing a preparation with $T_{\infty} = T_{H}$, we obtain
\begin{equation}
    Q_{\infty} \geq k_{B}T_{H}\ln 2.
\end{equation}
Thus, the gravitational factor is already incorporated into this expression; we must not multiply $T_H$ by an additional factor of $\sqrt{f(r_0)}$. Choosing $T_\infty=T_H$ fixes the temperature of the prepared stationary reservoir. It should not be taken to mean that the field surrounding an evaporating black hole occupies a global Gibbs state.

For a reservoir in Tolman equilibrium, the higher temperature measured closer to the horizon is offset by the redshift of the exchanged energy. The energy bound referred to infinity therefore remains unchanged when the device is moved at fixed $T_H>0$. The extremal limit is different, since $T_H$ itself approaches zero. Whether an actual reset becomes less costly cannot be decided from this temperature alone it depends on the reservoir levels, the memory Hamiltonian, and how the interaction is switched on and off. The horizon itself is not a location for a static device, and any limiting statement must account for this restriction.

\section{Heat and information balance}\label{sec:balance}

Throughout the following sections, $\mathcal{S}(\rho)=-\operatorname{Tr}(\rho\ln\rho)$ denotes dimensionless von Neumann entropy, and all logarithms are natural. The reservoir Hamiltonian $H_{R,\infty}$ is fixed during each protocol, and $T_\infty>0$ is its initial reference temperature; the final reservoir state need not be thermal. The background is held fixed during the process and backreaction is neglected. Let us first consider a memory $S$ and a reservoir $R$ in an initially uncorrelated state $\rho_{S}\otimes\rho_{R}$, such that the reservoir is initially prepared in the Gibbs state
\begin{equation}
 \rho_R = \frac{e^{-\beta_\infty H_{R,\infty}}}{Z}\ \text{with}\ \beta_{\infty} = \frac{1}{k_{B}T_{\infty}}.  
\end{equation}
After a global unitary evolution
\begin{equation}
    \rho'_{SR} =  U\rho_{SR}U^{\dagger},
\end{equation}
the energy received by the reservoir is given by
\begin{equation}\label{Q_infty}
    Q_{\infty} = \operatorname{Tr}[H_{R,\infty}(\rho'_{R}-\rho_{R})],
\end{equation}
where the state of reservoir is obtained $\rho_R=Tr_S\left[\rho_{SR}\right]$. The relative entropy is
\begin{equation}
    D(\rho_{R}'\vert\vert \rho_{R}) = \operatorname{Tr}[\rho'_{R}(\ln\rho'_{R}-\ln\rho_{R})],
    \label{Ecu:relative_entropy}
\end{equation}
and substituting the Gibbs state into Eq. (\ref{Ecu:relative_entropy}), we obtain
\begin{equation}
    D(\rho_{R}'\vert\vert \rho_{R}) = -\mathcal{S}(\rho'_{R}) +\beta_{\infty} \operatorname{Tr}(\rho'_{R}H_{R,\infty}) + \ln Z.
\end{equation}
Now, given that
\begin{equation}
    \mathcal{S}(\rho_{R}) = \beta_{\infty}\operatorname{Tr}(\rho_{R}H_{R,\infty})+\ln Z,
\end{equation}
results in
\begin{equation}
   \beta_{\infty} Q_{\infty} = \mathcal{S}(\rho'_{R}) - \mathcal{S}(\rho_{R}) + D(\rho_{R}'\vert\vert \rho_{R}). 
\end{equation}
On the other hand, the conservation of joint entropy under unitary evolution, together with the absence of initial correlations, allows one to write
\begin{equation}
    \mathcal{S}(\rho'_{R}) - \mathcal{S}(\rho_{R}) = \Delta \mathcal{S} + I(S':R'),
\end{equation}  
where 
\begin{equation}
    \Delta \mathcal{S} = \mathcal{S}(\rho_{S}) - \mathcal{S}(\rho'_{S}). 
\end{equation}
Therefore, 
\begin{equation}\label{reeb-wolf}
   \beta_{\infty} Q_{\infty} = \Delta \mathcal{S} + I(S':R') + D(\rho'_{R}\vert\vert \rho_{R}).
\end{equation}

The mutual information and relative entropy in Eq.~(\ref{reeb-wolf}) are non-negative, giving $\beta_\infty Q_\infty\geq\Delta\mathcal S$. Evaluating the excess heat requires a specified interaction and reservoir preparation, which we introduce below.

For a reservoir of finite physical Hilbert-space dimension $d\geq2$, Reeb and Wolf also establish the bound \cite{reeb2014improved}
\begin{equation}\label{finite_bound}
\beta_\infty Q_\infty\geq\Delta\mathcal S+
\frac{2(\Delta\mathcal S)^2}{\ln^2(d-1)+4},\qquad\Delta\mathcal S\geq0.
\end{equation}
For a maximally mixed qubit reset to $\operatorname{diag}(1-\epsilon,\epsilon)$, the entropy decrease is
\begin{equation}\label{binary_entropy}
\begin{aligned}
\Delta\mathcal S_\epsilon&=\ln2-h_2(\epsilon),\\
h_2(p)&=-p\ln p-(1-p)\ln(1-p),
\end{aligned}
\end{equation}
with natural logarithms and $0\ln0=0$. Substitution into Eq.~(\ref{finite_bound}) gives the approximate-erasure bound. At positive temperature, the initial joint state has full rank when the reservoir has finitely many finite-energy levels. Because a unitary cannot change this rank, the maximally mixed qubit cannot be reset to a pure state under these conditions. The case $\epsilon=0$ is therefore a useful ideal reference, rather than a finite-temperature outcome accessible in this model.

For a preparation at $T_\infty=T_H>0$, the geometry enters through Eq.~(\ref{T_H}). At fixed $d$, the temperature-dependent bound tends to zero as extremality is approached. This does not determine the heat of a particular protocol. Moreover, a bosonic cavity has unbounded occupations, its numerical cutoff is not a physical dimension to insert into Eq.~(\ref{finite_bound}). The cavity analysis therefore uses the information balance and an explicit spectral bound rather than identifying this cutoff with $d$.

\section{Heat transfer and external work}

The heat deposited in the reservoir need not equal the work supplied by the control apparatus. To distinguish them, consider a memory and reservoir outside the outer horizon with Hamiltonian
\begin{equation}
H_{\mathrm{tot},\infty}(t)=H_{S,\infty}(t)+H_{R,\infty}+H_{\mathrm{int},\infty}(t).
\end{equation}
All energies are referred to asymptotic time. The reservoir Hamiltonian is fixed, and the interaction vanishes at the initial and final times $t_i$ and $t_f$, respectively. Because the reservoir occupies an extended region, $H_{R,\infty}$ is defined with respect to the stationary time coordinate. A single redshift factor cannot generally convert the energies measured at different positions into this Hamiltonian.

Unitary evolution obeys
\begin{equation}\label{evolution}
i\hbar\dot\rho_{SR}=[H_{\mathrm{tot},\infty},\rho_{SR}].
\end{equation}
Differentiating $E_{\mathrm{tot},\infty}=\operatorname{Tr}(\rho_{SR}H_{\mathrm{tot},\infty})$, the term containing $\dot\rho_{SR}$ vanishes by cyclicity of the trace. Appendix~\ref{app:energy_conservation} gives the detailed derivation, first for constant coupling and then for the time-dependent pulse. The same cancellation holds in both cases; explicit time dependence of the Hamiltonian accounts for the work supplied by the control. Hence
\begin{equation}
\frac{dE_{\mathrm{tot},\infty}}{dt}
=\operatorname{Tr}(\rho_{SR}\partial_tH_{\mathrm{tot},\infty}).
\end{equation}
Integrating this expression and using the vanishing endpoint interaction gives
\begin{equation}\label{W_infty}
\begin{aligned}
W_\infty&=\int_{t_i}^{t_f}\operatorname{Tr}(\rho_{SR}\partial_tH_{\mathrm{tot},\infty})\,dt\\
&=\Delta E_{S,\infty}+Q_\infty,
\end{aligned}
\end{equation}
where $\Delta E_{S,\infty}=\operatorname{Tr}[\rho_S^fH_{S,\infty}(t_f)]-\operatorname{Tr}[\rho_S^iH_{S,\infty}(t_i)]$. A positive reservoir heat can therefore be supplied by a decrease in the memory energy, even when the external work vanishes.

At fixed reference temperature, we define $F_{S,\infty}=E_{S,\infty}-k_BT_\infty\mathcal S(\rho_S)$. Its change satisfies $\Delta F_{S,\infty}=\Delta E_{S,\infty}+k_BT_\infty\Delta\mathcal S$. Combining Eqs.~(\ref{reeb-wolf}) and (\ref{W_infty}) yields
\begin{equation}\label{dev}
W_\infty=\Delta F_{S,\infty}+k_BT_\infty(I_f+D_f)
\geq\Delta F_{S,\infty}.
\end{equation}
Here, $I_f=I(S':R')$ and $D_f=D(\rho_R^f\Vert\rho_R^i)$ for the initially thermal reservoir. The bound does not require positive work because $\Delta F_{S,\infty}$ can be negative. Under the finite-dimensional assumptions of Eq.~(\ref{finite_bound}), it strengthens to
\begin{equation}
W_\infty\geq\Delta F_{S,\infty}
+k_BT_\infty\frac{2(\Delta\mathcal S)^2}{\ln^2(d-1)+4}.
\end{equation}
Setting $T_\infty=T_H$ introduces the Bardeen temperature without changing the energy accounting. The behavior of $k_BT_\infty D_f$ in the cold limit is examined in Sec.~\ref{sec:cold}, after specifying the reservoir and its interaction.

\section{Scalar reservoir in an external cavity}

In this section, we aim to answer the question of what energies can be received by the field acting as a reservoir—a question that cannot be answered based solely on knowledge of the black hole's temperature. To do so, we must evaluate the energies associated with field excitations, which depend on the geometry and the boundary conditions imposed by the cavity. Thus, we seek to determine these energies by considering a scalar field confined between two spherical reflecting walls located outside the horizon and subsequently calculate the corresponding frequencies.

The reflecting walls are concentric with the black hole and placed symmetrically about $r_0$ in proper radial distance. Their positions are fixed by
\begin{equation}
    \rho(r) = \int_{r_{0}}^{r} \frac{dr'}{\sqrt{f(r')}},\ \rho(r_{a})=-\frac{L}{2},\ \rho(r_{b}) = \frac{L}{2},
\end{equation}
such that $r_{H}< r_{a}< r_{b}$. Note that this box does not cross the horizon.

The previous choice keeps the physical size of the cavity fixed, even if the geometry changes. We have considered this model in this section, taking as a direct precedent the work of Xu, Ong, and Yung \cite{xu2022landauer}, who studied—using perturbative methods—a qubit coupled to a field initially in the vacuum state or a thermal state.

Here, we consider a massless scalar field satisfying 
\begin{equation}\label{field}
    \square\, \Phi=\frac{1}{\sqrt{-\text{det} g}}\partial_{\mu}(\sqrt{-\text{det} g}\, g^{\mu\nu}\partial_{\nu}\Phi) = 0.
\end{equation}
{No mass term or additional curvature coupling is included. We also adopt the test-field approximation, the stress-energy of the scalar field is neglected in determining the background geometry.} 

Eq. (\ref{field}) is solved using the method of separation of variables by proposing a function of the form
\begin{equation}
    \Phi = e^{-i\omega t}u(r)Y_{\ell m}/r,
\end{equation}
from which it follows that
\begin{equation}
    \left[-\frac{d^{2}}{dr_{*}^{2}}+V_{\ell}(r)\right]u_{n\ell} = \frac{\omega_{n\ell}^{2}}{c^{2}}u_{n\ell},\  \frac{dr^{*}}{dr}=\frac{1}{f(r)}, 
\end{equation}
with
\begin{equation}
    V_{\ell} = f\left[\frac{\ell(\ell+1)}{r^{2}}+\frac{f'}{r}\right], 
\end{equation}
and we impose the Dirichlet boundary conditions 
\begin{equation}
    u_{n\ell}(r_{a}) = u_{n\ell}(r_{b})=0.
\end{equation}
Outside the outer horizon, the potential under consideration is non-negative; {indeed, $f>0$ and $f'\geq0$}, because $f'=2m r(r^{2}-2g^{2})/(r^{2}+g^{2})^{5/2}$. 

Now, using the integration by parts and Poincaré inequality gives \cite{kuznetsov2015sharp}
\begin{equation}
    \frac{\omega_{n\ell}^{2}}{c^{2}} = \frac{\int (|du/dr_*|^{2}+V_{\ell}|u|^{2})dr_{*}}{\int |u|^{2}dr_{*}}\geq \frac{\pi^{2}}{L_{*}^{2}},
\end{equation}
with $L_{*} = \int_{r_{a}}^{r_{b}}\frac{dr}{f(r)}$.

If the walls remain separated from the extremal horizon and $L_{*}$ has a finite limit, the fundamental frequency does not vanish even as $T_{H} \to 0$. This does not extend to walls that enter a throat of divergent optical length.

It is worth noting that a finite cavity does not necessarily possess a finite Hilbert dimension; indeed, bosonic occupancies are unbounded. The parameter $d$ in the aforementioned bounds is not identified with the number of modes or with a computational truncation. The thermal state of this cavity is a preparation; it is not automatically the Hawking radiation state.

We write $x=r_H/|g|$ and measure lengths in $|g|$ and times in $|g|/c$. We checked the cavity spectrum by solving the radial eigenvalue problem in proper distance. Using $u$ and $v=du/dr_*$ as integration variables makes it possible to impose the walls directly at their prescribed proper positions. For the $\ell=0$ sector used below, in units $|g|=c=\hbar=1$, their equations are
\begin{equation}
\frac{dr}{d\rho}=N,\qquad \frac{du}{d\rho}=\frac{v}{N},\qquad
\frac{dv}{d\rho}=\left(\frac{Nf'}{r}-\frac{\omega^2}{N}\right)u.
\end{equation}
The walls are located by integrating from $r(0)=3$ to $\rho=\pm0.2$. A fourth-order Runge--Kutta scheme with 3200 uniform intervals across the proper length is used. Starting with $u(-L/2)=0$ and $v(-L/2)=1$, each frequency is selected by the condition $u(L/2)=0$. We first bracket each root around $n\pi N_0/L$ and then carry out 42 bisection iterations. The same radial integration supplies $\int u^2\,d\rho/N$ and the overlap entering the coupling coefficient. These frequencies are roots of the discretized shooting equation; their accuracy with respect to the continuum problem must therefore be checked by refining the radial grid.

\subsection{Central redshift and cavity spectrum}

Consider a stationary clock located at $r_{0}$ that measures $\tau = N_{0}t$ with $N_{0} = \sqrt{f(r_{0})}$ and the frequency measured by this clock $\nu_{1}=\frac{\omega_{1}}{N_{0}}$. Outside the outer horizon of the Bardeen black hole $0<N_{0}<1$, the clock accumulates less proper time than the time elapsed at infinity.

For a cavity whose proper length is small compared to the local scales of geometric variation, in leading order, it has a fundamental frequency measured by the central clock of $\pi c/L$, and the corresponding frequency referred to the asymptotic time $t$ is $\pi c N_{0}/L$.

For the spherically symmetric mode $\ell = 0$, with Dirichlet boundary conditions at both walls and a cavity centred at $r_{0}$ in terms of the proper radial distance, the fundamental frequency admits the following expansion
\begin{align}\label{omega1}
    \omega_{1} = \frac{\pi c N_{0}}{L}\bigg[1 + \left(\frac{f'_{0}}{2\pi^{2}r_{0}} - \frac{1}{48}\left(\frac{f_{0}'^{\,2}}{f_{0}} - f''_{0}\right)\right)L^{2} + \mathcal{O}\left(\frac{L^{4}}{\mathcal{L}^{4}}\right)\bigg],
\end{align}
where $f_{0} = f(r_{0})$, and primes in this case denote derivatives with respect to $r$, but evaluated at $r_{0}$, and $\mathcal{L}$ represents a local scale of geometric variation. Indeed, $dr/d\rho = N$, $dN/d\rho =\frac{f'}{2}$ and $d^{2}N/d\rho^{2}=N f''/2$. By expanding $1/N$ and integrating symmetrically,
\begin{equation}
    L_{*} = \frac{L}{N_{0}}\left[1 + \frac{L^{2}}{48}\left(\frac{f_{0}'^{\,2}}{f_{0}} - f''_{0}\right)+ \cdots\right].
\end{equation}
Note that using $\omega_{1}^{2}/c^{2}\simeq (\pi/L_{*})^{2}+f_{0}f'_{0}/r_{0}$ yields Eq. (\ref{omega1}). The correction contains two contributions, the term proportional to $f'_{0}/r_{0}$ {arises from the radial potential}, whereas the remaining term accounts for the variation of the gravitational redshift across the cavity.

\begin{table}[tbp]
\caption{Fundamental frequency for $r_{0} = 3|g|$, $L=0.4|g|$.}
\begin{ruledtabular}\begin{tabular}{ccc}
$r_H/|g|$&$\omega_1|g|/c$&$\mathcal R$\\
2&3.550848982&0.999747881\\
1.6&3.951972775&0.999891188\\
$\sqrt2$&4.008810903&0.999907195
\end{tabular}\end{ruledtabular}\label{tab:spectrum}
\end{table}

The expansion of the short cavity is not uniform as the inner wall approaches the horizon—a point at which the assumption of a small geometric variation along the cavity may no longer hold. In Table~\ref{tab:spectrum}, the frequency—referenced to asymptotic time—increases by approximately 12.9\% as $x=r_H/|g|$ decreases from $2$ to its extremal value $\sqrt{2}$, whereas the dimensionless ratio $\mathcal{R}=\omega_{1} L/(\pi c N_{0})$ varies by only {0.0159\%}. Given that $\omega_{1} = (\pi c/L)N_{0}\mathcal{R}$ and $L$ is constant, this increase is primarily due to the variation in the central redshift factor $N_{0}$; the fundamental frequency, as measured by the central clock, remains virtually unchanged.

These results show that the vanishing of the Hawking temperature does not imply that the excitation energy of the confined field also vanishes. Furthermore, for the cavity configurations considered here, the proper length remains fixed, and the central redshift factor remains finite in the extremal state. Consequently, the fundamental excitation retains a non-zero energy—both locally, $\hbar\nu_1\simeq\pi\hbar c/L$, and with respect to infinity, $\hbar\omega_{1} = N_{0}\hbar\nu_{1}$. Thus, the thermal scale and the excitation gap describe distinct physical properties, the former characterizes thermal occupation, while the latter specifies the minimum energy required to excite the reservoir from its vacuum state. Whether or not information erasure necessitates such an excitation depends on the coupling between the memory and the reservoir, as well as the protocol employed, aspects that we will analyze next.

\section{Collective memory and protocol}

Once the frequencies supported by the cavity have been determined, this section will analyze how information is stored in a two-level quantum memory and how its interaction with the field allows for a reduction in the memory's entropy. To this end, we will specify the coupling between the two systems and calculate the energy transferred to the field, as well as the correlations generated during the process.

The fact that the memory has two levels (a ground state $|g\rangle$ and an excited state $|e\rangle$) does not imply that it interacts solely with the spherically symmetric modes $\ell=0$. At a fixed angular position $\Omega_0$, 
\begin{equation}
\sum_{m=-\ell}^{\ell} |Y_{\ell m}(\Omega_{0})|^{2}=(2\ell +1)/(4\pi).    
\end{equation}

For a point-like coupling at this angular position, this identity shows that sectors with $\ell>0$ are not eliminated by angular symmetry. We therefore adopt a collective two-level degree of freedom coupled to the angular average of the field, for which
\begin{equation}
\int\frac{d\Omega}{4\pi}Y_{\ell m} =  \frac{\delta_{\ell 0}\delta_{m0}}{\sqrt{4\pi}}.
\end{equation}

This angular average models a collective degree of freedom distributed over the spherical cavity rather than a detector localized at one angular position. It is a useful idealization because it isolates the radial field dynamics while keeping the memory--reservoir coupling explicit. We assume that the apparatus realizes this angular weighting; the model does not establish how to build such a device. In particular, its mode selection and the resulting quantitative predictions should not be transferred to a point-like detector, which generally couples to higher angular sectors.

Next, we introduce a radial profile with respect to proper distance as
\begin{equation}
 F_L(\rho)=\begin{cases}
 [1+\cos(2\pi\rho/L)]/L, &|\rho|\leq L/2,\\
 0, &|\rho|>L/2.\end{cases}
\end{equation}
This profile is normalized with respect to proper radial distance, $\int_{-L/2}^{L/2}F_L(\rho)\,d\rho=1$, and both the profile and its first derivative vanish at the walls. It specifies the radial weighting of the field in its coupling to the memory.

Using the central proper time $\tau = N_{0}t$ with $\nu_{n}=\omega_{n}/N_0$, we define the free and interaction Hamiltonians as follows
\begin{eqnarray}
  H_{0}  &=& \hbar\Omega |e\rangle\langle e| + \sum_{n} \hbar \nu_{n}a_{n}^{\dagger}a_{n},\label{Eq.H0}\\
  H_{int}(\tau) &=& \hbar\chi(\tau)\sigma_{x}\sum_{n}\kappa_{n}(a_{n}+a_{n}^{\dagger}).
  \label{Eq.HI}
\end{eqnarray}
Here, $|e\rangle$ is the excited state of the quantum memory, and $\sigma_{x} = |g\rangle\langle e|+|e\rangle\langle g|$ is the Pauli operator that induces transitions between the excited state and the ground state $|g\rangle$, which are separated by an energy gap of $\hbar\Omega$, and the dimensionless switching function $\chi$ controls how the interaction is turned on and off. It is also important to mention that $a_{n}^{\dagger}$ and $a_{n}$ are the creation and annihilation operators for excitations of the field's $n$-th mode, whose frequency relative to the central clock is $\nu_{n}$. The coefficients $\kappa_{n}$ characterize the coupling strength between the memory and each reservoir mode and have dimensions of angular frequency. Here, $n$ labels the radial modes in the selected $\ell=0$ sector. The coefficients $\kappa_n$ and switching function $\chi(\tau)$ are taken to be real, so the interaction Hamiltonian $H_{int}(\tau)$ is Hermitian.

The Hamiltonians in Eqs.~(\ref{Eq.H0}) and (\ref{Eq.HI}) are related to the Jaynes--Cummings model \cite{jaynes1963comparison}, which describes the coupling of a two-level system to a bosonic cavity mode. This model is widely used in quantum optics \cite{agarwal2012quantum} and in descriptions of superconducting qubits interacting with cavity fields \cite{clarke2008super,blais2021circuit}. Here the cavity serves as a finite-size reservoir.

In the multimode evolution of the combined memory-field system, the counter-rotating terms are retained. These are the terms $\sigma_+a_n^\dagger$ and $\sigma_-a_n$, which simultaneously excite or de-excite the memory and a field mode, respectively, where $\sigma_+=|e\rangle\langle g|$ and $\sigma_-=|g\rangle\langle e|$. In the weak-coupling regime, these rapidly oscillating terms can often be neglected within the rotating-wave approximation. They need not vanish, however, and can become important in the ultrastrong-coupling regime \cite{niemczyk2010circuit}, where the rotating-wave approximation may fail \cite{larson2021jaynes}. We therefore retain them in the multimode calculation.

To specify the overlaps between the field modes and the coupling profile, we set \(\hbar=c=1\), choose real radial eigenfunctions normalized such that \(\int u_n^2\,dr_*=1\), and write
\begin{equation}
    A_{n} = \frac{1}{\sqrt{8\pi\omega_{n}}}\int_{-L/2}^{L/2}F_{L}(\rho)\frac{u_{n}[r(\rho)]}{r(\rho)}d\rho,\qquad \kappa_{n} = \lambda A_{n}.
\end{equation}
Here, $A_n$ is the overlap, which measures the extent to which the spatial shape of the mode coincides with the profile perceived by the memory, and $\lambda$ sets the overall scale of the coupling in this convention. Thus, a single device can couple to different modes with varying intensities.

We use a pulse $\chi(\tau) = \sin^{2}(\pi\tau/\tau_{f})$ for $0<\tau<\tau_{f}$ and zero outside. The interaction vanishes at the endpoints. The memory starts in the maximally mixed state 
\begin{equation}
 \rho_{S}^{i}=\mathbb{I}_{S}/2=\frac{1}{2}|g\rangle\langle g|+{\frac{1}{2}|e\rangle\langle e|},   
\end{equation}
and its final error is characterized by $p_e=\langle e|\rho_S^f|e\rangle\leq\epsilon$. This implies a mixture with equal probability for the two states—not a coherent superposition of them—and there is no well-defined relative phase between the two possibilities within that mixture.

The target state is $|g\rangle$, so $p_e$ is the probability of finding the memory outside the target state after the interaction. This error corresponds to that specific initial state, not to the worst-case scenario across all inputs. In the following subsections, we consider an initially uncorrelated field prepared either in the vacuum or in a thermal state and specify each preparation separately.

\subsection{Resonant solution in vacuum}

To obtain an analytical reference for the erasure mechanism, we first restrict the reservoir to a single mode—initially in the vacuum state—and apply the rotating-wave approximation. This approximation is used solely for this reference model; the subsequent multimode calculation retains the counter-rotating terms. At resonance, $\Omega = \nu_1$, the Hamiltonian in the interaction picture is $H_I(\tau) = \hbar\kappa_1\chi(\tau)(\sigma_+ a + \sigma_- a^\dagger)$.

At exact resonance, the Hamiltonian in the interaction representation depends on time solely through the switching function $\chi(\tau)$; consequently, the relation $[H_{I}(\tau), H_{I}(\tau')] = 0$ holds, and the evolution is determined by the dimensionless pulse area
\begin{equation}\label{pulse_area}
\Theta=\kappa_1\int_0^{\tau_f}\chi(\tau)\,d\tau
=\kappa_1\int_0^{\tau_f}\sin^2\!\left(\frac{\pi\tau}{\tau_f}\right)d\tau
=\frac{\kappa_1\tau_f}{2}.
\end{equation}

The memory, initially in a state of maximal mixing, together with the field in the vacuum state, form the uncorrelated joint state
\begin{equation}
\rho_{SR}^{i} =\frac12|g,0\rangle\langle g,0| +\frac12|e,0\rangle\langle e,0|.
\end{equation}

The component $|g,0\rangle$ remains unaltered, since the field contains no quanta capable of exciting the memory, and the latter is already in its ground state. In contrast, the component $|e,0\rangle$ evolves within the single-excitation subspace
\begin{equation}
|e,0\rangle\longmapsto |\psi_f\rangle =\cos\Theta|e,0\rangle -i\sin\Theta|g,1\rangle.
\end{equation}
This state represents the vacuum Rabi oscillations \cite{agarwal1985vacuum}.
Therefore, the joint final state is:
\begin{equation}
\rho_{SR}^{f} =\frac12|g,0\rangle\langle g,0| +\frac12|\psi_f\rangle\langle\psi_f|.
\end{equation}

When taking the trace over the field, the terms proportional to $|0\rangle\langle1|$ and $|1\rangle\langle0|$ vanish, given that the vacuum and single-quantum states are orthogonal. The reduced state of the memory turns out to be
\begin{equation}
 \rho_S^f =\left(1-\frac12\cos^2\Theta\right)|g\rangle\langle g|  +\frac{1}{2}\cos^2\Theta\,|e\rangle\langle e|. 
\end{equation}
Therefore, the residual population in the excited state is $p_e=\frac12\cos^2\Theta$. For $0\leq\epsilon\leq1/2$, the chosen pulse shape, and fixed coupling $\kappa_1>0$, the shortest nonnegative duration satisfying $p_e=\epsilon$ is
\begin{equation}
\label{pulse_duration}\tau_f=\frac{2}{\kappa_1}\arccos\left(\sqrt{2\epsilon}\right).
\end{equation}

The entropy reduction is $\Delta\mathcal{S} =\ln2-h_{2}(\epsilon)$, with $h_2$ defined in Eq.~(\ref{binary_entropy}). The energy changes are
\begin{eqnarray}
    Q_{\infty} &=& \hbar\omega_{1}(1/2-\epsilon),\\
    \Delta E_{S,\infty} &=& -\hbar N_{0}\Omega(1/2-\epsilon),\ W_{\infty} = 0.
\end{eqnarray}
At resonance, $N_0\Omega=\omega_1$. Since the interaction vanishes at both endpoints, the energy balance gives $W_\infty=\Delta E_{S,\infty}+Q_\infty=0$. The field gains energy at the expense of the memory, rather than from the external control. The statement $W_\infty=0$ applies only to the switching pulse; it does not account for preparing the initial state or restoring the device afterward. The exact reset relies on the reservoir vacuum assumed in this example. A finite-dimensional reservoir prepared at positive temperature instead has full rank, and joint unitary evolution cannot produce the corresponding exact reset.

The final reservoir has probabilities $1/2+\epsilon$ and $1/2-\epsilon$ for zero and a quantum. Mutual information is
\begin{equation}
I_{f} = h_{2}(\epsilon) + h_{2}(1/2-\epsilon)-\ln2.
\end{equation}

For $\epsilon=0$, the memory becomes pure, while the reservoir acquires an entropy of $\ln 2$ and a mean energy of $\hbar\omega_1/2$. The final mutual information vanishes. Thus, the evolution transfers entropy to the reservoir without destroying global information; furthermore, perfect erasure entails a finite energy transfer supplied by the memory. A second-order perturbative approximation is insufficient on its own to establish a population exchange of order unity \cite{xu2022landauer}; the present resonant solution addresses the evolution non-perturbatively, within the framework of the rotating-wave approximation.

\subsection{Positive temperature: analytical reference}

Now we maintain the same single-mode resonant interaction and rotating wave approximation, but we prepare the reservoir in a thermal state,
\begin{equation}
    \tau_{R} = \sum_{n=0}^{\infty} p_{n} |n\rangle\langle n|,
\end{equation}
where $p_{n} = (1-q)q^{n}$ and $q=e^{-\beta_{\infty}\hbar\omega_{1}}$. Initially, the memory remains in a state of maximal mixing and is uncorrelated with the reservoir
\begin{equation}
    \rho_{SR}^{i} = \frac{\mathbb I_{S}}{2}\otimes \tau_{R}.
\end{equation}

The {oscillator matrix elements} introduce occupation-dependent coupling. In each excitation sector, the evolution is
\begin{equation}
    |e,n\rangle \to \cos(\Theta\sqrt{n+1}) |e,n\rangle - i \sin(\Theta\sqrt{n+1})|g,n+1\rangle,
\end{equation}
and for $n\geq 1$,
\begin{equation}
    |g,n\rangle \to \cos(\Theta\sqrt{n}) |g,n\rangle-i\sin(\Theta \sqrt{n}) |e,n-1\rangle.
\end{equation}

The state $|g,0\rangle$ {remains unchanged}. Thus, the initially excited component can emit a quantum, whereas an {initially} ground-state memory can absorb {a thermal quantum} and become excited. Such states are called the dressed states \cite{agarwal2012quantum} which are results of coupling of the two-level system and bosonic mode.

Averaging these processes over the {initial thermal} distribution gives
\begin{equation}
    p_{e} = \frac{1}{2}\sum_{n=0}^{\infty}p_{n}[\cos^{2}(\Theta\sqrt{n+1})+\sin^{2}(\Theta\sqrt{n})].
\end{equation}

Now, to obtain a lower bound, define
\begin{equation}
    C(\Theta) = \sum_{n=0}^{\infty} p_{n}\cos^{2}(\Theta\sqrt{n+1}).
\end{equation}
Using $p_{n+1} = q\,p_{n}$ and shifting the summation index,
\begin{equation}
    \sum_{n=0}^{\infty} p_{n}\sin^{2}(\Theta\sqrt{n}) = q(1-C(\Theta)).
\end{equation}
Therefore, 
\begin{equation}
    {p_e=\frac12\left[q+(1-q)C(\Theta)\right]} \geq \frac{q}{2}.
\end{equation}
For a target $p_{e}\leq \epsilon$, with $0<\epsilon<1/2$, a necessary condition is therefore
\begin{equation}
    \beta_{\infty}\hbar \omega_{1} \geq \ln\left(\frac{1}{2\epsilon}\right).
\end{equation}
This condition is not sufficient, the transition angles depend on the initial occupation via $\sqrt{n+1}$, such that a pulse that efficiently transfers one excitation sector does not necessarily transfer the others efficiently. In particular, under this protocol, perfect erasure at any strictly positive temperature is ruled out.

The final state of the reservoir is diagonal in the number basis,
\begin{equation}
    \rho_{R}^{f} = \sum_{n=0}^{\infty} p_{n}^{f}|n\rangle\langle n|,
\end{equation}
with
\begin{eqnarray}
    p_{n}^{f} &=& \frac{p_{n}}{2}[\cos^{2}(\Theta\sqrt{n})+\cos^{2}(\Theta\sqrt{n+1})]\nonumber\\
    && +\frac{p_{n-1}}{2}\sin^{2}(\Theta\sqrt{n})+\frac{p_{n+1}}{2}\sin^{2}(\Theta\sqrt{n+1}),
\end{eqnarray}
with $p_{-1}=0$. 

{Conservation of the total excitation number gives}
\begin{equation}
   Q_{\infty} = {\hbar}\omega_{1}\left(\frac{1}{2}-p_{e}\right).
\end{equation}
{The reduced memory state remains diagonal, so its entropy reduction is}
\begin{equation}
   \Delta\mathcal{S} = \ln 2 - h_{2}(p_{e}).
\end{equation}

Given that the joint evolution is unitary and the initial state is uncorrelated, the mutual information is
\begin{equation}
    I_{f} = h_{2}(p_{e}) + \mathcal{S}(\rho_{R}^{f})-\ln2-\mathcal{S}(\tau_{R}),
\end{equation}
where
\begin{equation}
    {\mathcal{S}(\rho_R^f)=-\sum}_{n} p_{n}^{f}\ln p_{n}^{f},\qquad \mathcal{S}(\tau_{R}) = -\sum_{n}p_{n}\ln p_{n}.
\end{equation}

The relative entropy with respect to the initial thermal state is
\begin{equation}
    D_{f} = \sum_{n}p_{n}^{f}\ln(p_{n}^{f}/p_{n}).
\end{equation}

Now, if we use $\ln p_{n} = \ln (1-q)-\beta_{\infty}\hbar\omega_1n$, we obtain
\begin{equation}
    D_{f} = \beta_{\infty}Q_{\infty} -[\mathcal{S}(\rho_{R}^{f})-\mathcal{S}(\tau_{R})].
\end{equation}
Combining this expression with mutual information yields the {Reeb--Wolf balance} (\ref{reeb-wolf}).
Therefore, thermal preparation affects both the achievable erasure error and the energy balance. The initially occupied field states allow for absorption processes that re-excite the memory, while the heat transferred to the reservoir includes contributions associated with the reduction of the memory's entropy, {the correlations generated}, and the reservoir's deviations from its thermal state.

\section{Spectral restriction of heat}

We now examine how the available reservoir energies constrain heat transfer when the temperature-dependent Landauer scale becomes small.

To this end, we assume a reservoir in its unique ground state $|0\rangle$ with zero ground-state energy. All other energies are at least $\Delta_{R,\infty} > 0$; this is written as
\begin{equation}
    H_{R,\infty} \geq \Delta_{R,\infty}(\mathbb I_{R}-|0\rangle\langle0|).
\end{equation}

The memory is initially in a maximally mixed state and uncorrelated with the reservoir, so that
\begin{equation}
\rho_{SR}^{i} =\frac{\mathbb I_{S}}{2}\otimes |0\rangle\langle0|.
\end{equation}
Here, we assume a unitary evolution of the combined memory-reservoir system, without any additional entropy sink; the two non-zero eigenvalues of the initial state are both $1/2$, and the unitary evolution preserves them. Consequently, the final state satisfies
\begin{equation}
\rho_{SR}^{f}\leq\frac12\mathbb I_{SR}.  
\end{equation}
In particular, $\langle g,0|\rho_{SR}^{f}|g,0\rangle\leq\frac12$. Furthermore, the probability of finding the memory in an excited state and the reservoir in its ground state cannot exceed the total population in the excited state
\begin{equation}
\langle e,0|\rho_{SR}^{f}|e,0\rangle \leq p_e\leq\epsilon.
\end{equation}
By adding these contributions together, we get
\begin{eqnarray}
P_R(0) &=& \langle g,0|\rho_{SR}^{f}|g,0\rangle+ \langle e,0|\rho_{SR}^{f}|e,0\rangle\nonumber\\ 
& \leq& \frac12+\epsilon. 
\end{eqnarray}
Therefore, the final energy of the reservoir satisfies
\begin{equation}
U_{R,\infty}^{f} \geq\Delta_{R,\infty}[1-P_R(0)] \geq\Delta_{R,\infty}\left(\frac12-\epsilon\right).
\end{equation}
Given that the initial energy of the reservoir is zero, $Q_{\infty} = U_{R,\infty}^{f}$, and consequently
\begin{equation}
Q_\infty\geq\Delta_{R,\infty}\left(\frac12-\epsilon\right), \qquad 0\leq\epsilon\leq\frac12.
\end{equation}

The reservoir must acquire a nonzero probability of excitation as the memory approaches its target pure state. Since every excited reservoir state has energy at least $\Delta_{R,\infty}$, the minimum excitation probability $1/2-\epsilon$ implies a minimum mean energy transfer. This argument does not require that the transferred energy originate exclusively from the memory; external work may also contribute.
The resonant protocol saturates this bound when the quantum used has energy $\Delta_{R,\infty}$. The bound depends on the initial state and the allowed resources; it is not an additional universal law to be added to Landauer's principle. Spectral erasure restrictions and optimal erasure protocols have precedents in \cite{mohammady2016minimising}. We do not claim a general optimality theorem for our pulse or a new universal Landauer inequality. Here the relevance of the bound is geometric: the gap is determined by the field spectrum of a finite exterior cavity in the Bardeen spacetime. Its persistence at extremality applies to this specified reservoir and does not imply a nonzero cost for every conceivable erasure protocol.

For an initially thermal reservoir with ground-state population $p_0$, retaining the initially maximally mixed and uncorrelated memory, the largest joint eigenvalue is $p_0/2$. The same argument gives $P_R(0)\leq p_0/2+\epsilon$, and hence
\begin{equation}
U_{R,\infty}^{f}\geq\Delta_{R,\infty}\left(1-\frac{p_0}{2}-\epsilon\right).
\end{equation}
Subtracting the initial reservoir energy $U_{R,\infty}^{i}$ yields
\begin{equation}
Q_\infty\geq\Delta_{R,\infty}\left(1-\frac{p_0}{2}-\epsilon\right)-U_{R,\infty}^{i}.
\end{equation}
This lower bound can become negative and then does not, by itself, establish a positive minimum heat transfer. The thermal and spectral bounds must both hold when their respective assumptions apply; they cannot simply be added.

\section{Multimode model results}

Here, we examine whether the erasure protocol calibrated within the single-mode resonant approximation remains effective when multiple cavity modes and counter-rotating terms are taken into account. We use $|g|=c=\hbar=1$, with energy unit $E_*=\hbar c/|g|$, and we fix the cavity at the center $r_{0}=3$ {and keep the proper length} $L=0.4$. The memory and the coupling are calibrated at $x = r_{H}/|g|=2$, where
$\Omega = 7.852001494,\ \kappa_{1} = 0.01, \lambda = 0.2221395902$.
For the reference target $\epsilon = 10^{-3}$, Eq.~(\ref{pulse_duration}) gives
$\tau_f=305.2120093$. The initial memory state is $\mathbb I_{S}/2$, and the retained field modes are initially in their joint vacuum. Each evolution is performed on a {fixed, static geometry}. When comparing different values of $x$, we retain the same memory {frequency} $\Omega$, pulse duration $\tau_{f}$, overall coupling parameter $\lambda$, and profile $F_{L}(\rho)$. The cavity {frequencies} and mode overlaps are recalculated for each geometry. Therefore, both the detunings
\begin{equation}
    \delta_{n} = \nu_{n}-\Omega,
\end{equation}
and the {individual} couplings $\kappa_{n} = \lambda A_{n}$ can change. This comparison tests the response of the specified device to different geometries; it does not describe a time-dependent geometry during a single erasure operation.

For $M$ retained radial modes, the numerical reservoir basis {consists} of states 
\begin{equation}
    |n_{1},\dots,n_{M}\rangle,\qquad \sum_{j=1}^{M} n_{j}\leq K,  
\end{equation}
where $K$ is a cutoff on the {total occupation}, rather than a separate {cutoff} for each mode.

The reservoir basis has dimension
\begin{equation}
    d_{R} = \binom{M+K}{K},
\end{equation}
and the joint memory and reservoir basis has dimension $2d_{R}$. The cutoff value defines the numerical approximation and must be verified to ensure convergence.

The evolution includes the full interaction Hamiltonian, including its counter-rotating terms. Given that the interaction vanishes at the beginning and end of the pulse, the final energy balance can be expressed using the energies of the free system and the reservoir. {The heat received by the field}, referenced to infinity, is
\begin{eqnarray}
    Q_{\infty} = \sum_{n}\hbar\omega_{n}(\langle a_{n}^{\dagger}a_{n}\rangle_{f}-\langle a_{n}^{\dagger}a_{n}\rangle_{i}),
\end{eqnarray}
and the work supplied through the switching is
\begin{equation}
    W_{\infty} = N_{0}\int_{0}^{\tau_{f}}\operatorname{Tr}(\rho_{SR}\partial_{\tau}H_{int})d\tau.
\end{equation}
These quantities satisfy
\begin{equation}
 W_{\infty} = \Delta E_{S,\infty} + Q_{\infty} \qquad \Delta E_{S,\infty} = N_{0}\hbar\Omega \left(p_{e}-\frac{1}{2}\right).   
\end{equation}
The factor $N_{0}$ converts the energy defined via the proper-time Hamiltonian into energy referred to infinity; the field contribution already employs the {corresponding asymptotic frequencies $\omega_n$}.

For the numerical checks, we evolve the pure-state components of the initial mixture in the interaction picture and combine them with their statistical weights. In the basis $|s,\mathbf n\rangle$, with $s=0,1$, the free angular-frequency eigenvalues are $e_{s,\mathbf n}=s\Omega+\sum_n\nu_n n_n$. The amplitudes satisfy
\begin{equation}
i\dot c_a(\tau)=\chi(\tau)\sum_b V_{ab}e^{i(e_a-e_b)\tau}c_b(\tau),
\end{equation}
where $V_{ab}$ connects opposite memory states and field occupations differing by one quantum, with matrix elements $\kappa_n\sqrt{n_n+1}$ for creation and $\kappa_n\sqrt{n_n}$ for annihilation. Both rotating and counter-rotating transitions are retained within the truncated basis. For vacuum input, the initial vectors $|g,\mathbf0\rangle$ and $|e,\mathbf0\rangle$ are propagated separately and combined with weights $1/2$. A fixed-step fourth-order Runge--Kutta method is used with $\Delta\tau=\tau_f/N_t$ and the step counts shown in Table~\ref{tab:conv}. The work integral is accumulated at the same Runge--Kutta stages. We monitor the norm of each propagated state before renormalization, since its drift provides a check on the integration. The reduced density matrices are then obtained from the ensemble weighted by $1/2$, taking the appropriate partial traces and retaining the off-diagonal entries.

We tested temporal convergence at $x=1.6$ with $M=5$ modes and $K=3$. Doubling $N_t$ from 80000 to 160000 changes $p_e$ by less than $10^{-12}$ and $Q_\infty/E_*$ by less than $4\times10^{-12}$. Over the same comparison, the largest norm error before renormalization decreases from about $2.42\times10^{-12}$ to $1.18\times10^{-13}$. Because the radial frequencies are unchanged, this test isolates time stepping and says nothing about the remaining spatial error. The pulse vanishes at both endpoints, allowing us to obtain the work independently from the energy balance, $W_\infty=Q_\infty+N_0\hbar\Omega(p_e-1/2)$. Thus, for the same geometry and coupling, the quoted time-step differences imply
\begin{equation}
 \frac{|\delta W_\infty|}{E_*}<4\times10^{-12}+\frac{N_0\hbar\Omega}{E_*}\,10^{-12}.
\end{equation}
This is a check of time-step sensitivity rather than a complete estimate of numerical uncertainty. We therefore repeated the calculation with an independent spatial discretization and time integrator. The radial operator was discretized by Chebyshev collocation on the proper-distance interval, imposing Dirichlet conditions at both walls. The map $r(\rho)$ was obtained by inversion of the proper-distance integral with 64-point Gauss--Legendre quadrature; normalization and overlaps used 200-point quadrature. Increasing the polynomial degree from 72 to 96 changed the first eleven eigenfrequencies by less than $3.0\times10^{-13}$ relatively for the three geometries considered.

The independent evolution uses the embedded Dormand--Prince 5(4) method, with local componentwise error scale $\eta[1+\max(|c_a^{\rm old}|,|c_a^{\rm new}|)]$, and $\eta=10^{-12}$ unless specified otherwise. The switching work is integrated alongside the amplitudes, and no final normalization is imposed. This integrator is distinct from DOP853. Table~\ref{tab:independent-cutoffs} compares five and seven modes at extremality with the same spatial and temporal settings.

An earlier shooting calculation gave a fundamental frequency $4.008810902269$, compared with $4.008810902542$ in the collocation calculation. Replacing only this frequency in the latter evolution changes $p_e$ by approximately $2.3\times10^{-9}$ and $Q_\infty/E_*$ by $9.1\times10^{-9}$. This controlled test largely accounts for the difference between the earlier implementations, the long pulse makes the result sensitive to small spectral shifts. Agreement under changes of the occupation cutoff does not, on its own, quantify the error arising from the cavity spectrum.

\begin{table}[tbp]
\caption{Independent extremal evolution using Chebyshev collocation of degree 72 and Dormand--Prince 5(4) with $\eta=10^{-12}$. Extra digits resolve changes between cutoffs and are not absolute error estimates.}\label{tab:independent-cutoffs}
\begin{ruledtabular}\begin{tabular}{rrrr}
$M$ & $K$ & $Q_\infty/E_*$ & $W_\infty/E_*$\\
5 & 3 & 1.989703491417 & 0.000317017737 \\
7 & 3 & 1.989703491331 & 0.000317017737 \\
\end{tabular}\end{ruledtabular}
\end{table}

As a separate check, we varied the radial grid while leaving the quantum-evolution solver unchanged. At extremality, the lowest frequency and its coupling were calculated on three proper-distance grids with the same shooting condition and normalization. The values in Table~\ref{tab:radial-refinement} indicate that $\kappa_1$ is insensitive to this refinement at the reported precision. This check is restricted to the radial calculation; it does not assess errors from higher modes, the time integrator, or the occupation cutoff. The separate cutoff comparison is given in Table~\ref{tab:independent-cutoffs}.
\begin{table}[tbp]
\caption{Independent refinement of the extremal cavity's fundamental radial eigenfrequency and coupling, using the shooting method. $N_\rho$ denotes the number of proper-distance grid points used to tabulate the geometry and potential.}
\label{tab:radial-refinement}
\begin{ruledtabular}
\begin{tabular}{rrr}
$N_\rho$ & $\omega_1 |g|/c$ & $\kappa_1 |g|/c$\\
1001 & 4.008810898065 & 0.010001296035850\\
2001 & 4.008810901428 & 0.010001296035905\\
4001 & 4.008810902269 & 0.010001296035910
\end{tabular}
\end{ruledtabular}
\end{table}

\begin{table}[tbp]\caption{Residual excited-state population of the memory for three- and five-mode reservoirs, including counter-rotating terms and a total occupation cutoff of two quanta.}\label{tab:multi}
\begin{ruledtabular}
\begin{tabular}{ccc}
$x$ & Three modes & Five modes\\
2 & 0.00100015846 & 0.00100015845\\
1.6 & 0.00315509540 & 0.00315509592\\
$\sqrt2$ & 0.00366741113 & 0.00366741170
\end{tabular}
\end{ruledtabular}
\end{table}

Table \ref{tab:multi} shows that the residual population in the excited state increases as the geometry approaches extremality. For five modes, this value rises from approximately $0.00100016$ at $x=2$ to $0.00366741$ at $x=\sqrt{2}$. Consequently, the fixed pulse results in less complete erasure near extremality, although the final population in the excited state remains small. This change is accompanied by variations in the mode frequencies and couplings; the comparison does not isolate their individual contributions.

The calibration target $\epsilon = 10^{-3}$ corresponds to the rotating-wave reference in the fundamental mode. For the calibration geometry, the {full multimode result} exceeds this target by approximately ${1.58\times 10^{-7}}$. This difference quantifies the correction for the selected parameters, rather than establishing an exact saturation of the target.

\begin{table}[tbp]
\caption{Convergence checks for the residual population in excited states in the extremal limit. $M$ is the number of radial modes retained, $K$ is the upper limit for total occupancy, and \textit{steps} indicates the number of time-integration steps.}\label{tab:conv}
\begin{ruledtabular}
\begin{tabular}{rrrr}
Modes & $K$ & steps & $p_e$\\
3&2&80000&0.0036674111327\\
5&2&80000&0.0036674117047\\
7&2&100000&0.0036674117270\\
9&2&100000&0.0036674117290\\
11&2&100000&0.0036674117293\\
5&3&160000&0.0036674116669
\end{tabular}
\end{ruledtabular}
\end{table}

The comparisons in Table \ref{tab:conv} show variations of less than $10^{-7}$ in the reported final population. With a fixed value of $K=2$ and 100000 integration steps, increasing the number of modes from seven to eleven results in a variation of approximately $2.3\times 10^{-12}$ in $p_{e}$. Comparing five modes between $K=2$ and $K=3$ yields a variation of approximately $3.8\times 10^{-11}$; however, the integration resolution also changes in this comparison. Consequently, this assesses the combined effect of modifying the occupancy threshold and the temporal resolution, rather than isolating each of these sources of numerical error. The displayed digits are numerical outputs rather than a uniform error estimate. Table~\ref{tab:independent-cutoffs} provides a separate comparison for heat and work. Their changes are small over the cuts tested, although this is not an absolute error bound. The information terms $I_f$ and $D_f$ are checked independently below; small differences between implementations of the radial spectrum still constrain the meaningful precision of all these numbers.

The detunings, overlap integrals, and duration of the pulse all enter the reset probability. Accordingly, the decreasing fidelity in this sequence is a result for the apparatus with those parameters held fixed. A different memory gap or pulse shape could change the outcome, but has not been evaluated here. The restriction to collective angular coupling discussed above also remains part of this interpretation.

\section{Heat decomposition in the Reeb--Wolf identity}

We evaluate the information balance for $x=1.6$, five radial modes, and total occupation cutoff $K=3$. The energy scale is $E_*=\hbar c/|g|$, and the preparation temperature is specified by
\begin{equation}
B=\beta_\infty\hbar\omega_1=\frac{\hbar\omega_1}{k_BT_\infty}.
\end{equation}

The reference used for the tabulated numerical decomposition is specifically the Gibbs state conditioned on the total-occupation subspace
\begin{equation}
\begin{aligned}
\tau_{R,K}&=\frac{P_K e^{-\beta_\infty H_{R,\infty}}P_K}{Z_K},\\
Z_K&=\sum_{\sum n_j\leq K}e^{-B\sum_j(\nu_j/\nu_1)n_j}.
\end{aligned}
\end{equation}
Here, $P_K$ projects onto $\sum_j n_j\leq K$. In the reported finite-basis evaluation of $D_f$ and $D_i$, $\tau_R$ means $\tau_{R,K}$; this is not the unconditioned product Gibbs state used in the mode-correlation identity below. Within this basis, the propagated initial configurations are the union of those whose normalized probability exceeds $10^{-14}$ for at least one of $B=12,16,24$. The selected probabilities are renormalized separately for each preparation, and both memory states are evolved for every selected field configuration.

Here, the quantity $B$ compares the fundamental excitation energy with the thermal energy scale. For a fixed geometry, increasing $B$ is equivalent to decreasing the preparation temperature of the reservoir.

The truncated reservoir Hilbert space has dimension $d_R=\binom{8}{3}=56$, while the joint memory-reservoir Hilbert space has dimension $112$. This numerical cutoff is not interpreted as the physical dimension of the reservoir in the dimension-dependent Reeb-Wolf inequalities.

We use $\Delta\mathcal S$ as defined in Sec.~\ref{sec:balance}, with $I_f=\mathcal S(\rho_S^f)+\mathcal S(\rho_R^f)-\mathcal S(\rho_{SR}^f)$ and $D_f=D(\rho_R^f\Vert\tau_{R,K})$. For an exact Gibbs initial state, Eq.~(\ref{reeb-wolf}) applies without an initial correction. For the retained and renormalized initial distribution actually used in these tables, the corresponding balance is
\begin{equation}\label{general_balance}
\beta_\infty Q_\infty=\Delta\mathcal S+I_f+D_f-D_i,
\end{equation}
since the initial memory and reservoir are uncorrelated. Here $D_i=D(\rho_R^i\Vert\tau_{R,K})$.

To compare these dimensionless terms with the reservoir heat, we multiply them by $k_BT_\infty$. The three positive contributions in the tables sum to $Q_\infty+k_BT_\infty D_i$, rather than exactly to $Q_\infty$. The magnitude of $D_i$ is reported below. Its effect is below the precision shown in the tables, but retaining it distinguishes the actual truncated preparation from an ideal Gibbs state. The entropy reduction, mutual information, and relative entropy enter one balance equation rather than representing separate transfers of heat.

\begin{table*}[tp]\caption{Entropy reduction, final mutual information, and relative entropy of the reservoir for three preparation temperatures at $x=1.6$. The calculation considers five radial modes with a total occupancy limit of $K=3$. All entropies use natural logarithms.}\label{tab:rw}
\begin{ruledtabular}
\begin{tabular}{rrrrr}
$B$ & $p_e$ & $\Delta\mathcal S$ & $I_f$ & $D_f$\\
12 & 0.00315908455 & 0.671804747 & 0.021270943 & 5.269015296\\
16 & 0.00315516894 & 0.671827281 & 0.021298750 & 7.256391266\\ 
24 & 0.00315509591 & 0.671827702 & 0.021299569 & 11.231150428
\end{tabular}
\end{ruledtabular}
\end{table*}
In Table \ref{tab:rw}, it can be observed that the final population in memory, the reduction in entropy, and the mutual information vary only slightly across the temperature range considered. Conversely, $D_{f}$ increases significantly with $B$. To express each dimensionless contribution in energy units, it must be multiplied by $k_BT_\infty$.

Using $E_{*} = \hbar c/|g|$, the conversion is
\begin{equation}
    \frac{k_{B}T_{\infty}}{E_{*}} = \frac{\omega_{1}|g|/c}{B}.
\end{equation}
Note that at $x=1.6$, the calculated frequency is $\omega_{1} |g|/c \simeq 3.951972775$. The resulting energy contributions are presented in Table \ref{tab:rwenergy}.

\begin{table*}[tp]
\caption{Energy contributions to the Reeb-Wolf balance for the configurations in Table \ref{tab:rw}, expressed in units of $E_{*} = \hbar c/|g|$.}\label{tab:rwenergy}
\begin{ruledtabular}
\begin{tabular}{rrrrr}
$B$ & $k_BT_\infty\Delta\mathcal S$ & $k_BT_\infty I_f$ & $k_BT_\infty D_f$ & $Q_\infty$\\
12 & 0.221246173 & 0.007005182 & 1.735250417 & 1.963501771\\
16 & 0.165940195 & 0.005260755 & 1.792316295 & 1.963517246\\
24 & 0.110626866 & 0.003507305 & 1.849383363 & 1.963517534
\end{tabular}
\end{ruledtabular}
\end{table*}

The total heat remains close to $1.9635 E_{*}$, while the contributions $k_{B}T_{\infty}\Delta\mathcal{S}$ and $k_{B}T_{\infty} I_{f}$ decrease as the preparation temperature is lowered. The contribution from relative entropy increases and accounts for the majority of the energy transferred. Therefore, a larger value of $D_{f}$ does not imply a divergent heat; its energetic contribution is given by the product $k_BT_\infty D_f$.

The reduced density matrix of the reservoir is constructed by including its off-diagonal elements. Its von Neumann entropy is evaluated from its eigenvalues, rather than being based solely on occupation probabilities. This distinction is important, given that disregarding coherences can alter both the entropy of the reservoir and its relative entropy.

For the initial selection described above, discarding thermal energy eigenstates and renormalizing their probabilities gives
\begin{equation}\label{initial_relative}
D_i=-\ln P_{\mathrm{ret}},
\end{equation}
where $P_{\mathrm{ret}}$ is evaluated within the same reference $\tau_{R,K}$. This correction concerns discarded initial weights, not errors from the mode or dynamical occupation cutoffs.

Re-evaluating these initial weights with 85-digit decimal arithmetic, using the stored mode frequencies, gives the following corrections. The extra precision avoids cancellation in $1-P_{\mathrm{ret}}$; it does not imply 85-digit accuracy of the spectrum.
\begin{table}[tbp]

\caption{Initial relative entropy caused by omission of small initial weights within the $M=5$, $K=3$ Gibbs reference. Energy corrections are expressed in units of $E_*=\hbar c/|g|$.}\label{tab:initialcorrection}
\begin{ruledtabular}\begin{tabular}{ccc}
$B$ & $D_i$ & $k_BT_\infty D_i/E_*$\\
12 & $7.01353\times10^{-16}$ & $2.30977\times10^{-16}$\\
16 & $4.32060\times10^{-21}$ & $1.06718\times10^{-21}$\\
24 & $1.63971\times10^{-31}$ & $2.70004\times10^{-32}$
\end{tabular}\end{ruledtabular}
\end{table}
As Table~\ref{tab:initialcorrection} shows, the energy correction is below $2.4\times10^{-16}E_*$, far below the displayed precision of Table~\ref{tab:rwenergy}. If standard floating-point arithmetic returns zero for this tiny correction, that reflects limited machine precision rather than an exact cancellation.

Relative to the unconditioned product Gibbs state of the same five modes, one must additionally include $\ln(Z_{\mathrm{prod}}/Z_K)$, where $Z_{\mathrm{prod}}=\prod_{j=1}^5[1-e^{-B\nu_j/\nu_1}]^{-1}$. This additional term is approximately $1.42517\times10^{-21}$, $1.60381\times10^{-28}$, and $2.03109\times10^{-42}$ for $B=12,16,24$, respectively. The estimates apply only to the five modes included in the reference partition function. They say nothing about the error associated with modes left out of the calculation or the later cutoff imposed during the time evolution.

A small residual in the balance does not constitute independent evidence of the evolution, given that relative entropy is itself calculated from energy and entropy.

\paragraph{Independent numerical cross-check.}
The collocation and Dormand--Prince calculation reproduces the rounded entries in Tables~\ref{tab:rw} and \ref{tab:rwenergy}. At $x=1.6$, tightening $\eta$ from $10^{-11}$ to $10^{-12}$ changes $p_e$ by less than $1.2\times10^{-13}$, $Q_\infty/E_*$ by $2.4\times10^{-12}$, $I_f$ by $1.8\times10^{-12}$, and $D_f$ by $1.5\times10^{-11}$. At the tighter tolerance, the propagated vectors have orthonormality errors below $6\times10^{-14}$ and the independently integrated work agrees with the endpoint energy balance within $8.3\times10^{-14}E_*$. Joint entropy is also evaluated from the propagated ensemble rather than assumed to be conserved. Small negative reservoir eigenvalues of order $10^{-16}$ are consistent with floating-point roundoff.

Increasing $M$ from five to seven, while keeping $K=3$ fixed, results in changes of less than $2.0 \times 10^{-11}$ in the value of $p_e$, $7.7 \times 10^{-11}$ in $Q_{\infty}/E_{*}$, $1.2 \times 10^{-10}$ in $I_f$, and $4.7 \times 10^{-10}$ in $D_f$ across the three configurations. Subsequently, increasing $K$ from three to four, with $M=5$ held constant, produces changes of less than $2.3 \times 10^{-14}$ in $Q_\infty/E_*$ and $1.3 \times 10^{-13}$ in $D_f$ values close to the numerical rounding error. Since the variations for each parameter were assessed separately, these differences should not be interpreted as a bound for the combined numerical error. Furthermore, the independent evolution of the vacuum in a single mode allows for the recovery of the target analytical value of $p_e = 10^{-3}$ with a precision of $4 \times 10^{-15}$.

A preparation at Hawking temperature would correspond to
\begin{equation}
    B_{H} = \frac{\hbar \omega_{1}}{k_{B}T_{H}}.
\end{equation}
We obtain $B_{H}\simeq 223.1064$ for $x=2$ and $B_{H}\simeq 505.1330$ for $x=1.6$—values that are significantly larger than the values $B = 12, 16, 24$ considered here. Consequently, these tables describe reservoir preparations that are hotter than the corresponding Hawking temperature. They test the information balance of an externally prepared reservoir, not the dynamics of a field in a Hawking-temperature Gibbs state. At the quoted Hawking values of $B_H$, the thermal populations would be much smaller; the present tables do not establish the corresponding dynamical or numerical results. Varying $B$ changes the reservoir preparation at fixed geometry and does not represent an evaporation trajectory.

\subsection{Cold limit and internal correlations}\label{sec:cold}

Let us consider the limit $T_{\infty}\to 0^{+}$, keeping the geometry fixed and assuming a specific protocol. This limit modifies the preparation of the reservoir and is distinct from the {black-hole extremal limit}—a process that can also alter mode frequencies and couplings.

For an initially maximally mixed two-level memory, we have
\begin{equation}
 0\leq \Delta \mathcal{S}\leq \ln 2,  \qquad  0\leq I_{f}\leq 2\ln2.   
\end{equation}
Therefore,
\begin{equation}
    k_{B}T_{\infty}\Delta\mathcal{S}\to 0,\ \text{and}\ k_{B}T_{\infty}I_{f}\to 0.
\end{equation}
These limits stem from the finite dimension of the memory and do not require the final state of the reservoir to be thermal.

The relative entropy contribution behaves differently. We define the reservoir free energy at the preparation temperature by
\begin{equation}
    F_{R,\infty}(\rho) =  \operatorname{Tr}(\rho H_{R,\infty})-k_{B}T_{\infty}\mathcal{S}(\rho). 
\end{equation}
For the Gibbs reference $\tau_{R}$, we have
\begin{equation}
  k_{B}T_{\infty}D(\rho_{R}^{f}||\tau_{R}) = F_{R,\infty}(\rho_{R}^{f})-F_{R,\infty}(\tau_{R}).  
\end{equation}
For a reservoir initially prepared in the exact thermal state $\tau_R$, this identity can be written as
\begin{equation}
k_{B}T_{\infty}D_{f} =  Q_{\infty} -k_{B}T_{\infty}[\mathcal{S}(\rho_{R}^{f})-\mathcal{S}(\tau_{R})].    
\end{equation}
If the heat approaches a finite value $Q_\infty^{(0)}$ when \(T_\infty\to0^+\), and $k_{B}T_{\infty}[\mathcal{S}(\rho_{R}^{f})-\mathcal{S}(\tau_{R})]\to 0$, then
\begin{eqnarray}
    \lim_{T_{\infty}\to 0^{+}} k_{B}T_{\infty}D_{f}=Q_\infty^{(0)}.
\end{eqnarray}
This limit is positive if the protocol leaves a non-zero mean excitation energy in the reservoir. The existence of a spectral gap, by itself, does not determine this energy; it is also necessary to specify the final excitation probabilities.

For a truncated initial preparation, the corresponding identity involves $D_f-D_i$ and the actual initial reservoir entropy. Recovering the same limit for $k_BT_\infty D_f$ additionally requires $k_BT_\infty D_i\to0$.

Thus, a finite heat transfer can persist even though the entropy-reduction and mutual-information contributions, expressed in energy units, both vanish. At exactly zero temperature, the reference state is the vacuum. For any reservoir state with nonzero support outside the vacuum, $D(\rho_R^f\Vert|0\rangle\langle0|)$ is infinite. The finite energy contribution must therefore be obtained as a limit from positive temperatures, rather than by directly evaluating $0\times\infty$.

For a multimode reservoir with a product thermal reference,
\begin{equation}
    \tau_{R} = \bigotimes_{n} \tau_{n}, 
\end{equation}
the relative entropy further decomposes as
\begin{equation}
    D(\rho_{R}^{f}||\bigotimes_{n} \tau_{n}) = \sum_{n}D(\rho_{n}^{f}||\tau_{n}) +\mathcal{I}_{modes},
\end{equation}
where $\rho_n^f$ is the reduced state of mode \(n\), and
\begin{equation}
    \mathcal{I}_{modes} = \sum_{n}\mathcal{S}(\rho_{n}^{f})-\mathcal{S}(\rho_{R}^{f}) = D(\rho_{R}^{f}||\bigotimes_{n}\rho_{n}^{f})\geq 0.
\end{equation}

This quantity measures the total correlations between the reservoir modes, including both classical and quantum correlations. It should be noted that it differs from $I_{f}$, which measures the correlations between the memory and the reservoir as a whole. Therefore, summing only the relative entropies of the individual modes omits the contribution of the correlations generated within the reservoir.

Here, the product reference is the thermal state of the independent modes before conditioning on a total occupation cutoff. Conditioning on the total-occupation cutoff and renormalizing can introduce correlations between the mode occupations, even though the untruncated thermal reference factorizes.

\section{Scalar Hawking radiation near extremality}

The preceding calculation measures the energy deposited in a cavity by a controlled memory reset. Hawking emission poses a different question, how much scalar-field energy reaches infinity without reflecting walls? We therefore remove the cavity boundaries and impose scattering conditions on the radial modes. Both calculations use the Bardeen background, but their field states and boundary conditions differ. In particular, a smaller outgoing Hawking flux cannot be used to infer a smaller excitation energy for the confined reservoir.

For a massless scalar field minimally coupled to the static background, the Hawking power measured at infinity is
\begin{equation}
    P_{\infty} = \frac{1}{2\pi}\sum_{\ell = 0}^{\infty}(2\ell +1)\int_{0}^{\infty}\frac{\hbar\omega\Gamma_{\ell}(\omega)}{e^{\hbar\omega/(k_{B}T_{H})}-1}d\omega.
\end{equation}
Here, $\omega$ is the frequency with respect to asymptotic time, and $\Gamma_{\ell}(\omega)$ is the transmission probability through the external curvature potential for an angular sector $\ell$. The factor $2\ell +1$ accounts for the angular degeneracy. These transmission probabilities—also referred to as greybody factors—modify the thermal spectrum that reaches infinity. Specifically, the calculations for the Bardeen geometry have been presented in the work of Konoplya et al.~\cite{konoplya2023bardeen}. Scalar absorption by the Bardeen black hole was also investigated directly in Ref.~\cite{macedo2014absorption}; thus, we do not present the scattering potential or the greybody-factor computation by itself as a new result. The additional question considered here is how the near-extremal emission scale compares with the independently specified energetic requirements of an exterior erasure device.

{The radiation formula includes all angular sectors, unlike the collective cavity coupling used above. The Unruh state describes outgoing Hawking radiation without an incoming thermal bath at infinity and is not a global Gibbs state of the exterior field. Knowledge of $T_H$ alone therefore does not establish the thermal-reservoir assumptions required for the Reeb--Wolf identity.}

At low frequency, the absorption cross-section is related to the transmission factors by
\begin{equation}
    \sigma_{abs}(\omega) =  \frac{\pi c^{2}}{\omega^{2}}\sum_{\ell = 0}^{\infty} (2\ell +1)\Gamma_{\ell}(\omega). 
\end{equation}
For the minimally coupled massless scalar, the low-frequency absorption result \cite{das1997universality} gives 
\begin{equation}
    \lim_{\omega\to0}\sigma_{\mathrm{abs}}(\omega)=4\pi r_H^2.
\end{equation}
Since the $\ell=0$ sector dominates in this limit, it follows that
\begin{equation}
     \Gamma_{0}(\omega) \simeq \frac{4r_{H}^{2}\omega^{2}}{c^{2}}.
\end{equation}
Thus, 
\begin{equation}
    P_{\infty} \simeq \frac{2\hbar r_{H}^{2}}{\pi c^{2}}\int_{0}^{\infty}\frac{\omega^{3}}{e^{\hbar\omega/(k_{B}T_{H})}-1}d\omega.
\end{equation}
Introducing $z=\hbar\omega/(k_BT_H)$ and using
\begin{equation}
\int_0^\infty\frac{z^3}{e^z-1}\,dz=\frac{\pi^4}{15},
\end{equation}
we obtain
\begin{equation}
P_\infty\simeq\frac{2\pi^3r_H^2}{15\hbar^3c^2}(k_BT_H)^4.
\end{equation}
This approximation requires the low-frequency transmission formula to describe the frequencies that contribute significantly to the thermal integral. Its normalization is derived from the absorption cross-section.

Now, in order to examine the approach to extremality, we define $x = \frac{r_{H}}{|g|}$ and $y = r_{H}/|g| - \sqrt{2}$. Near the extremal point (for small positive $y$), we have
\begin{align}
\label{near_temperature}\frac{k_{B}T_{H}|g|}{\hbar c} &= \frac{y}{6\pi}+\mathcal{O}(y^2),\\
\label{near_mass}m-m_{e} &= |g|[\sqrt{3}y^{2}/4+\mathcal{O}(y^{3})],\\
\overline{P} &\equiv \frac{P_{\infty}g^{2}}{\hbar c^{2}}\simeq \frac{2\pi^{3}}{15}x^{2}\left(\frac{k_{B}T_{H}|g|}{\hbar c} \right)^{4}\approx \frac{y^{4}}{4860\pi}.\label{near_power}
\end{align}

The numerical results support a quartic decrease of the scalar luminosity as the black hole approaches extremality. This agrees with the leading low-frequency approximation, in which the Hawking temperature vanishes linearly while the horizon area remains finite. Nevertheless, the analytical coefficient requires care, the low-frequency absorption result is derived at fixed nonextremal geometry, whereas here the low-frequency and extremal limits are approached together. The calculations support the behavior along the sequence studied, but do not establish a uniform approximation at exact extremality.

We checked the radiated power by integrating the radial wave equation with an ingoing condition near the horizon and matching the solution to travelling waves at large radius. In dimensionless units, write $v=du/dr_*$, so that $du/dr=v/f$ and $dv/dr=(V_\ell-\omega^2)u/f$. For the nonextremal configurations, the baseline integration starts just outside the outer horizon, at $r_{\mathrm{in}}=x+10^{-8}(x-\sqrt2)$, with $u=1$ and $v=-i\omega$. This prescription is used only for $x>\sqrt{2}$, rather than at exact extremality. At large radius, matching to $u=A_{\mathrm{in}}e^{-i\omega r_*}+A_{\mathrm{out}}e^{i\omega r_*}$ gives
\begin{equation}
\Gamma_\ell=|A_{\mathrm{in}}|^{-2},\qquad
|A_{\mathrm{in}}|^2=\frac14|u+iv/\omega|^2.
\end{equation}
For the tabulated powers, we integrate in $\ln[(r-x)/(x-\sqrt2)]$ using adaptive Dormand--Prince 5(4), with componentwise tolerance scale $10^{-11}[1+\max(|U_a^{\rm old}|,|U_a^{\rm new}|)]$ for $U=(\operatorname{Re}u,\operatorname{Im}u,\operatorname{Re}(v/\omega),\operatorname{Im}(v/\omega))$. The metric function is evaluated in a form that avoids subtracting nearly equal terms near the horizon. For the values reported below, we place the inner boundary at an offset $10^{-9}(x-\sqrt2)$ from the horizon and match at $\max(100x,240/\omega)$. The frequency integral uses 96-point Gauss--Legendre quadrature over $0<z<40$, where $z=\hbar\omega/(k_BT_H)$. We sum partial waves up to $\ell=4$ for $x=2$ and $1.6$. In the four near-extremal cases we use $\ell\leq1$; separate checks extending to $\ell=3$ indicate that higher partial waves do not change the quoted precision.

Independent fixed-step Runge--Kutta integrations and refined Simpson quadratures were also compared. We halved the radial step limits and decreased the frequency spacing from $0.125$ to $0.03125$. The comparison also cautions against interpreting agreement at a single quadrature resolution as validation of every reported digit. We report five significant figures for the powers in Table~\ref{tab:power}; ratios and slopes are calculated from the unrounded results. The last column approaches the coefficient predicted by the low-frequency expansion.

To check the power-law exponent independently of its coefficient, we evaluate the slope between neighboring data points
\begin{equation}
 p_{\mathrm{eff}}(y_1,y_2)=\frac{\ln[\overline P(y_2)/\overline P(y_1)]}{\ln(y_2/y_1)}.
\end{equation}
For the successive pairs $(0.02,0.01)$, $(0.01,0.005)$ and $(0.005,0.001)$, the unrounded values give $p_{\mathrm{eff}}\simeq3.99225$, $3.99589$ and $3.99854$, respectively. This trend provides a useful numerical check of the quartic power, while leaving the uniformity of the near-extremal scattering approximation as a separate question.

\begin{table}[tbp]
\caption{Dimensionless Hawking power of a minimally coupled massless scalar field for selected Bardeen configurations. Here, $x=r_H/|g|$, $y=x-\sqrt{2}$, and $\overline P=P_\infty g^2/(\hbar c^2)$.}\label{tab:power}
\begin{ruledtabular}
\begin{tabular}{crr}
Configuration & $\overline P$ & $\overline P/[y^4/(4860\pi)]$\\
$x=2$ & $2.6018\times10^{-6}$ & ---\\
$x=1.6$ & $6.9418\times10^{-8}$ & ---\\
$y=0.02$ & $1.0363\times10^{-11}$ & 0.988910\\
$y=0.01$ & $6.5118\times10^{-13}$ & 0.994233\\
$y=0.005$ & $4.0815\times10^{-14}$ & 0.997069\\
$y=0.001$ & $6.5457\times10^{-17}$ & 0.999413
\end{tabular}
\end{ruledtabular}
\end{table}

Table~\ref{tab:power_convergence} summarizes single-parameter checks against a baseline with tolerance $10^{-10}$, horizon offset $10^{-8}(x-\sqrt2)$, outer radius $\max(100x,120/\omega)$, and 64 frequency nodes on $0<z<32$. Across the six configurations, tightening the tolerance to $10^{-11}$ changes the power by less than $3.8\times10^{-9}$ relatively. Moving the inner boundary ten times closer changes it by less than $4.5\times10^{-11}$, and extending the frequency range to $z=40$ with 96 nodes changes it by less than $2\times10^{-10}$. Within the scattering calculation adopted here, the variations are smaller than the precision quoted for the luminosities. They should not be read as a rigorous bound on the combined numerical error, particularly in the exact-extremal limit.

\begin{table}[tbp]
\caption{Relative luminosity changes on increasing frequency nodes from 64 to 96 and doubling the outer matching scale from $120/\omega$ to $240/\omega$, with other baseline settings fixed. The last column is the baseline fractional $\ell=1$ contribution.}\label{tab:power_convergence}
\begin{ruledtabular}\begin{tabular}{cccc}
$y$ & Frequency change & Matching change & $\ell=1$ fraction\\
0.020 & $2.80\times10^{-13}$ & $1.35\times10^{-9}$ & $8.88\times10^{-10}$ \\
0.005 & $2.17\times10^{-13}$ & $1.93\times10^{-9}$ & $1.56\times10^{-12}$ \\
0.001 & $2.22\times10^{-13}$ & $2.17\times10^{-9}$ & $9.56\times10^{-16}$ \\
\end{tabular}\end{ruledtabular}
\end{table}

When transmission is very small, subtracting the large, nearly equal incident and reflected fluxes is ill-conditioned. This affects higher partial waves and can also occur for $\ell=0$ at very low frequency. For example, at $y=0.001$ the largest such residual is about $2\times10^{-4}$ at the low-frequency end of the final integration. It must not be interpreted as a relative error estimate for the integrated luminosity. We instead assess its stability through the separate integration, boundary, and quadrature tests above.

The suppression of Hawking luminosity near extremality concerns the spontaneous emission of the unconfined field. This does not imply that excitations within the externally confined reservoir become energetically free. The finite cavity gap follows from its reflecting boundaries, whereas the luminosity is computed for outgoing modes in the unconfined geometry.

\section{Conditional evaporation and equivalent rate}

We next ask what time dependence would follow if the calculated luminosity were assigned to mass loss along the static Bardeen family, keeping $g$ fixed. This quasistatic prescription is expressed as
\begin{equation}
    c^{2}\frac{d\mathcal{M}}{dt}= -P_{\infty}.
\end{equation}
This formulation does not constitute a dynamical solution to the coupled equations of gravity and matter. In particular, the conventional formulation based on nonlinear electrodynamics includes the parameter $s=|g|/(2m)$ in its Lagrangian \cite{ayon2000bardeen, ma2014corrected}. Thus, varying $m$ while keeping $g$ constant alters this parameter; consequently, it cannot be automatically interpreted as an evolution within a theory of matter in which the couplings remain fixed.

Writing Eq.~(\ref{mass_m}) in terms of $x$ gives
\begin{equation}
    m(x) = |g|\frac{(x^{2}+1)^{3/2}}{2x^{2}},\qquad \mathcal{M} = \frac{c^{2}m}{G}, 
\end{equation}
thus
\begin{equation}
    \frac{dm}{dx} = \frac{|g|}{2}\frac{\sqrt{x^{2}+1}(x^{2}-2)}{x^{3}}.
\end{equation}

Combining this derivative with the assumed energy-loss equation yields
\begin{equation}
    \frac{dx}{dt} = - \frac{2GP_{\infty}}{c^{4}|g|}\frac{x^{3}}{\sqrt{x^{2}+1}(x^{2}-2)}.
\end{equation}
Near extremality, inserting Eqs.~(\ref{near_mass}) and (\ref{near_power}) into the energy-loss equation gives
\begin{equation}
    \frac{c^{4}|g|}{G}\frac{\sqrt{3}}{2}y\frac{dy}{dt} \simeq -\frac{\hbar c^{2}}{g^{2}}\frac{y^{4}}{4860\pi}.
\end{equation}
Now, introducing the dimensionless time 
\begin{equation}
 \mathcal{T} = c\ell_{P}^{2}t/|g|^{3},   
\end{equation}
we obtain
\begin{equation}
    \frac{dy}{d\mathcal{T}} \simeq -\frac{y^{3}}{2430\sqrt{3}\pi}.
\end{equation}
Integrating the leading-order equation from $y_{0} = y(\mathcal{T}_{0})>0$ gives
\begin{equation}
   y(\mathcal{T}) \simeq \left[ y_{0}^{-2}+\frac{\mathcal{T}-\mathcal{T}_{0}}{1215\sqrt{3}\pi} \right]^{-1/2}.
\end{equation}
Within this prescription, extremality is approached asymptotically rather than reached in finite time. This does not establish the existence of a stable extremal remnant or determine the endpoint of physical evaporation. The formal late-time scalings are $y$, $T_{H}\propto t^{-1/2}$, $\mathcal{M}-\mathcal{M}_{e}\propto t^{-1}$ and $P_{\infty}\propto t^{-2}$.

These results inherit the assumptions of both the quasi-static prescription and the near-extremal luminosity approximation. To evaluate a limitation of the description of continuous energy loss, let us consider the energy available above extremality as
\begin{equation}
    E_{exc} = (\mathcal{M}-\mathcal{M}_{e})c^{2}\simeq \frac{c^{4}|g|}{G}\frac{\sqrt{3}}{4}y^{2}.
\end{equation}
A characteristic emitted quantum has energy
\begin{equation}
    E_{\gamma}\sim \alpha k_{B}T_{H},
\end{equation}
where \(\alpha\) is a dimensionless factor determined by the emission spectrum. Their ratio is
\begin{equation}
    \frac{E_{\gamma}}{E_{exc}}\sim \frac{2\alpha}{3\pi \sqrt{3}}\frac{\ell_{P}^{2}}{g^{2}y}.
\end{equation}
Requiring an individual characteristic quantum to carry only a small fraction of the available energy gives
\begin{equation}
    y \gg \frac{2\alpha}{3\pi\sqrt{3}}\frac{\ell_{P}^{2}}{g^{2}}.
\end{equation}

For a finite factor $\alpha$ of order unity, this condition is expressed parametrically as $y\gg \ell_P^2/g^2$. The inequality is required if the residual energy loss is to be approximated as continuous. It is only one consistency check, however, and cannot establish the validity of the entire semiclassical evolution.

Finally, for $T_{H}>0$, we define the equivalent thermal-energy rate 
\begin{equation}
    \dot{N}_{eq} = \frac{P_{\infty}}{k_{B}T_{H}\ln 2} \simeq \frac{c}{|g|}\frac{y^{3}}{810\ln 2}.
\end{equation}
The rate is simply a way of measuring radiated power relative to the thermal Landauer energy per bit. It should not be interpreted as the number of bits erased in an implemented protocol. Combining the restricted identity $T_H\,dS_{BH}=\mathcal Cc^2\,d\mathcal M$ with the assumed energy-loss equation gives
\begin{equation}
    -\frac{\dot{S}_{BH}}{k_{B}\ln 2} = \mathcal{C} \dot{N}_{eq},
\end{equation}
with $\mathcal C=\frac{r_H^3}{(r_H^2+g^2)^{3/2}}$.
Therefore, even the decrease in the geometric entropy of the horizon—expressed in units of $k_{B}\ln 2$—generally differs from the equivalent thermal energy rate. Neither of these quantities, on its own, identifies the number of bits erased via a physical protocol. At exact extremality, $P_\infty/(k_BT_H\ln2)$ is undefined. Within the adopted near-extremal approximation, its limit is zero because $P_\infty\propto y^4$ whereas $T_H\propto y$.

\section{Conclusions}
\label{sec:conclusions}

We have examined information erasure in the Bardeen geometry by distinguishing the thermal Landauer scale from the energy transferred in an explicit memory--reservoir protocol. The Hawking temperature and the scale $k_B T_H\ln 2$ vanish at extremality, whereas the geometric horizon entropy remains finite. This behavior does not determine the operational cost of erasure. For the exterior cavity considered here, the walls remain separated from the horizon and the field retains a nonzero excitation gap.

For an initially maximally mixed memory and a reservoir (cavity) in its unique ground state, unitary evolution and the reservoir gap imply $Q_\infty\geq\Delta_{R,\infty}(1/2-\epsilon)$. The resonant single-mode protocol saturates this bound when the exchanged quantum has energy $\Delta_{R,\infty}$. Its vanishing external work follows from the decrease in the memory energy, which supplies the heat received by the field. In the full multimode calculation, the fixed pulse produces a small increase in the residual excited-state population as extremality is approached. This result characterizes the specified device and initial memory state; it does not establish a universal erasure cost or an optimal protocol.

Thermal preparation affects both the achievable erasure error and the energy balance. Initially occupied field states permit absorption processes that re-excite the memory. The Reeb--Wolf identity relates the reservoir heat to the memory entropy decrease, the correlations generated between memory and reservoir, and the reservoir's departure from its initial thermal state. These terms constitute an information balance for the same energy transfer. In the cold limit, the entropy reduction and mutual information contributions vanish in energy units, while $k_B T_\infty D_f$ can remain finite when the protocol deposits a finite excitation energy in the reservoir.

The numerical thermal preparations studied here are hotter than the corresponding Hawking temperature. They test the information balance of an externally prepared cavity reservoir and do not establish the dynamics of a field prepared at $T_\infty=T_H$. Varying the preparation parameter $B$ at fixed geometry changes the initial reservoir state and does not represent an evaporation trajectory.

Separately, the unconfined scalar-field calculation supports a quartic suppression of the near-extremal luminosity, $P_\infty\propto y^4$. Under the assumed quasistatic evolution at fixed $g$, this gives $y\propto t^{-1/2}$ and an asymptotic approach to extremality. These results remain conditional on the scattering approximation and the adopted energy-loss prescription; they do not establish a stable remnant or the endpoint of physical evaporation. In particular, the suppression of spontaneous Hawking radiation does not imply that the confined reservoir can receive excitations at vanishing energy. The radiative luminosity and the heat transferred during erasure characterize different processes.

These findings return us to the question motivating this work: could information be erased without an energetic cost if its environment were cooled to absolute zero? The extremal Bardeen geometry provides a limiting setting in which to examine this possibility, since its Hawking temperature, referred to infinity, vanishes while its horizon area remains finite. However, the disappearance of the thermal Landauer scale does not establish that a physical erasure operation becomes energetically free. For the reservoir and resources considered here, resetting the memory requires a nonzero probability of exciting the field, whose spectral gap remains finite. Consequently, a finite heat transfer persists even in the vacuum preparation. This energy can be supplied by the memory itself, so its persistence does not imply a strictly positive external work requirement. The result shows that a vanishing temperature alone is insufficient to determine the energetic cost of erasure.

For future information technologies, including quantum devices, this distinction suggests that cooling alone cannot guarantee arbitrarily small erasure costs. Cooling suppresses thermal occupations and the associated re-excitation of the memory, but the energy exchanged during a reset also depends on the accessible spectrum, the target error, and the control protocol. Assessing the cost of repeated operation additionally requires accounting for the preparation and restoration of these resources. We have not shown that every conceivable information-processing task must dissipate a nonzero amount of energy, and different reservoirs or control strategies may perform better. The narrower conclusion supported by our calculation is that the energetic cost of this reset cannot be inferred from the thermal Landauer scale alone, its spectrum, initial state, and control procedure have to be specified.

\subsection*{Declaration of competing interest}

The authors declare that they have no known competing financial
interests.

\subsection*{Conflict of Interest}

The authors declare that they do not have any conflict of interest.

\subsection*{Data Availability Statement}

No datasets were generated, analyzed or used in this study.




\appendix
\section{Energy conservation and switching work}
\label{app:energy_conservation}

To clarify the energy accounting when counter-rotating terms are retained, we consider the total memory--reservoir Hamiltonian
\begin{equation}
    H(\tau)=H_0+\chi(\tau)V,
\end{equation}
where $H_0$ describes the uncoupled memory and reservoir, $V$ is the interaction operator, and $\chi(\tau)$ controls the coupling. The Hamiltonian is defined with respect to the central proper time $\tau$. The background and the free Hamiltonian $H_0$ are fixed throughout the operation. We first establish conservation for constant coupling, and then include the explicit time dependence of the switching pulse used in the main text.

We first assume a constant coupling, $\chi(\tau)=\chi_0$, so that
\begin{equation}
    H=H_0+H_{\mathrm{int}},
    \qquad H_{\mathrm{int}}=\chi_0V.
\end{equation}
Since $H_0$ and $V$ have no explicit time dependence, this gives
\begin{equation}
    \frac{\partial H}{\partial\tau}=0.
\end{equation}
However, the joint state $\rho_{SR}(\tau)$ may still evolve. Thus, the time independence of the Hamiltonian does not imply that the density operator is constant.

The total mean energy is
\begin{equation}
    E_{\mathrm{tot}}(\tau)
    =\operatorname{Tr}\!\left[\rho_{SR}(\tau)H\right].
\end{equation}
In a fixed orthonormal basis $\{|j\rangle\}$ of the joint Hilbert space, we can write
\begin{equation}
    E_{\mathrm{tot}}(\tau)
    =\sum_j\langle j|\rho_{SR}(\tau)H|j\rangle.
\end{equation}
Differentiating with respect to $\tau$ yields
\begin{equation}
\begin{aligned}
    \frac{dE_{\mathrm{tot}}}{d\tau}
    &=\sum_j\left\langle j\left|
    \frac{d}{d\tau}(\rho_{SR}H)
    \right|j\right\rangle\\
    &=\operatorname{Tr}\!\left[
    \frac{d}{d\tau}(\rho_{SR}H)
    \right].
\end{aligned}
\end{equation}

Applying the product rule while preserving the order of the operators,
\begin{equation}
    \frac{d}{d\tau}(\rho_{SR}H)
    =\dot{\rho}_{SR}H
    +\rho_{SR}\frac{\partial H}{\partial\tau},
\end{equation}
where the dot denotes differentiation with respect to $\tau$. Therefore,
\begin{equation}
    \frac{dE_{\mathrm{tot}}}{d\tau}
    =\operatorname{Tr}(\dot{\rho}_{SR}H)
    +\operatorname{Tr}\!\left(
    \rho_{SR}\frac{\partial H}{\partial\tau}
    \right).
\end{equation}
The second term vanishes under the assumption of a time-independent Hamiltonian, leaving
\begin{equation}\label{app:energy_derivative}
    \frac{dE_{\mathrm{tot}}}{d\tau}
    =\operatorname{Tr}(\dot{\rho}_{SR}H).
\end{equation}

The joint unitary evolution obeys the von Neumann equation,
\begin{equation}
    i\hbar\dot{\rho}_{SR}=[H,\rho_{SR}].
\end{equation}
Solving for the time derivative of the density operator and expanding the commutator gives
\begin{equation}
\begin{aligned}
    \dot{\rho}_{SR}
    &=-\frac{i}{\hbar}(H\rho_{SR}-\rho_{SR}H).
\end{aligned}
\end{equation}
Substituting into Eq.~(\ref{app:energy_derivative}), we obtain
\begin{equation}
    \frac{dE_{\mathrm{tot}}}{d\tau}
    =-\frac{i}{\hbar}
    \operatorname{Tr}\!\left[
    (H\rho_{SR}-\rho_{SR}H)H
    \right].
\end{equation}
Expanding the product inside the trace gives
\begin{equation}
    \frac{dE_{\mathrm{tot}}}{d\tau}
    =-\frac{i}{\hbar}
    \operatorname{Tr}\!\left(
    H\rho_{SR}H-\rho_{SR}H^2
    \right),
\end{equation}
and then, this becomes
\begin{equation}
    \frac{dE_{\mathrm{tot}}}{d\tau}
    =-\frac{i}{\hbar}
    \left[
    \operatorname{Tr}(H\rho_{SR}H)
    -\operatorname{Tr}(\rho_{SR}H^2)
    \right].
\end{equation}

The cyclic property of the trace implies
\begin{equation}
    \operatorname{Tr}(H\rho_{SR}H)
    =\operatorname{Tr}(\rho_{SR}HH)
    =\operatorname{Tr}(\rho_{SR}H^2).
\end{equation}
This equality does not require $H$ and $\rho_{SR}$ to commute. Consequently,
\begin{equation}
\begin{aligned}
    \frac{dE_{\mathrm{tot}}}{d\tau}
    &=-\frac{i}{\hbar}
    \left[
    \operatorname{Tr}(\rho_{SR}H^2)
    -\operatorname{Tr}(\rho_{SR}H^2)
    \right]\\
    &=0.
\end{aligned}
\end{equation}
Hence, the total mean energy is conserved, expressed by
\begin{equation}\label{app:total_energy_conservation}
    E_{\mathrm{tot}}(\tau)=E_{\mathrm{tot}}(0).
\end{equation}
This is conservation of the total energy, including the interaction energy. It does not imply that $\langle H_0\rangle$ is separately conserved. For constant coupling,
\begin{equation}
    \frac{d}{d\tau}\langle H_0\rangle
    =-\frac{d}{d\tau}\langle H_{\mathrm{int}}\rangle.
\end{equation}
Thus, changes in the free energy expectation can be compensated by changes in the interaction energy. Here ``free'' refers to the uncoupled Hamiltonian, not to thermodynamic free energy. Retaining counter-rotating terms does not invalidate the conservation law.

For the actual pulse, $\chi(\tau)$ varies in time. The commutator contribution to $dE_{\mathrm{tot}}/d\tau$ still vanishes by the same trace identity, but the explicit derivative of the Hamiltonian must be retained as
\begin{equation}\label{app:switching_power}
\begin{aligned}
    \frac{dE_{\mathrm{tot}}}{d\tau}
    &=\operatorname{Tr}\!\left(\rho_{SR}\frac{\partial H}{\partial\tau}\right)\\
    &=\dot\chi(\tau)\operatorname{Tr}(\rho_{SR}V).
\end{aligned}
\end{equation}
This is the power exchanged with the prescribed control, measured with respect to the central clock. A time-dependent coupling allows nonzero work, although its net value can vanish for a particular protocol.

In the static geometry, $\tau=N_0t$ with constant $N_0$, so the Hamiltonian generating asymptotic-time evolution is $H_{\mathrm{tot},\infty}=N_0H$. The switching work referred to infinity is therefore
\begin{equation}\label{app:switching_work}
    W_\infty=N_0\int_0^{\tau_f}
    \dot\chi(\tau)\operatorname{Tr}(\rho_{SR}V)\,d\tau.
\end{equation}
This factor converts between the two time generators; it is not a common local redshift assigned to each part of the extended reservoir. Since the pulse satisfies $\chi(0)=\chi(\tau_f)=0$, the endpoint interaction energies vanish and Eq.~(\ref{app:switching_work}) gives
\begin{equation}
    W_\infty=\Delta E_{S,\infty}+Q_\infty,
\end{equation}
in agreement with Eq.~(\ref{W_infty}). The constant-coupling result and the pulsed energy balance are therefore consistent. The zero work found for the resonant rotating-wave reference is a property of that protocol, not a general consequence of unitary evolution.
\end{document}